\documentclass[british,english,showpacs,preprintnumbers,amssymb,aps,notitlepage,twocolumn]{revtex4-1}
\usepackage{lmodern}

\usepackage[T1]{fontenc}
\usepackage[latin9]{inputenc}
\usepackage{color}
\usepackage{float}
\usepackage{amsmath}
\usepackage{graphicx}

\makeatletter

\providecommand{\tabularnewline}{\\}

\usepackage[dvipsnames]{xcolor}
\usepackage{amsfonts,amsmath} 

\usepackage{caption}
\usepackage{etoolbox} 
\patchcmd{\thebibliography}{\chapter*}{\section*}{}{}

\usepackage[breaklinks]{hyperref}
\hypersetup{
    colorlinks,
    linkcolor={red},
    citecolor={magenta},
    urlcolor={blue}
}

\usepackage[paper=a4paper,top=1.1in,bottom=1.2in,right=0.68in,left=0.68in]{geometry} 

\g@addto@macro\UrlBreaks{\do\-}
\definecolor{LightGray}{rgb}{0.8,0.8,0.8}

\@ifundefined{showcaptionsetup}{}{%
 \PassOptionsToPackage{caption=false}{subfig}}
\usepackage{subfig}
\makeatother

\usepackage{babel}
\begin{document}

\title{Pegasus: The second connectivity graph for large-scale quantum annealing
hardware}

\author{Nike Dattani}
\email{n.dattani@cfa.harvard.edu}

\affiliation{Harvard-Smithsonian Center for Astrophysics}

\affiliation{National Research Council of Canada}

\author{\vspace{-2mm}
Szilard Szalay}
\email{szalay.szilard@wigner.mta.hu}

\affiliation{Wigner Research Centre for Physics}

\author{\vspace{-2mm}
Nicholas Chancellor}
\email{nicholas.chancellor@durham.ac.uk}

\affiliation{Joint Quantum Centre, Durham University}
\begin{abstract}
Pegasus is a graph which offers substantially increased connectivity
between the qubits of quantum annealing hardware compared to the graph
Chimera. It is the first fundamental change in the connectivity graph
of quantum annealers built by D-Wave since Chimera was introduced
in 2009 and then used in 2011 for D-Wave's first commercial quantum
annealer. In this article we describe an algorithm which defines the
connectivity of Pegasus and we provide what we believe to be the best
way to graphically visualize Pegasus in order to see which qubits
couple to each other. As supplemental material, we provide a wide
variety of different visualizations of Pegasus which expose different
properties of the graph in different ways. We provide an open source
code for generating the many depictions of Pegasus that we show.
\end{abstract}
\maketitle
\begin{figure*}

\caption{Three different depictions of the Chimera graph, with open edges to
show that the pattern repeats.\label{fig:chimera}}

\subfloat[Tilted classic]{

\includegraphics[width=0.33\textwidth]{metapostfigs/fig_ChimeraStdRect5x5_crop}}\subfloat[Diamond]{

\includegraphics[width=0.33\textwidth]{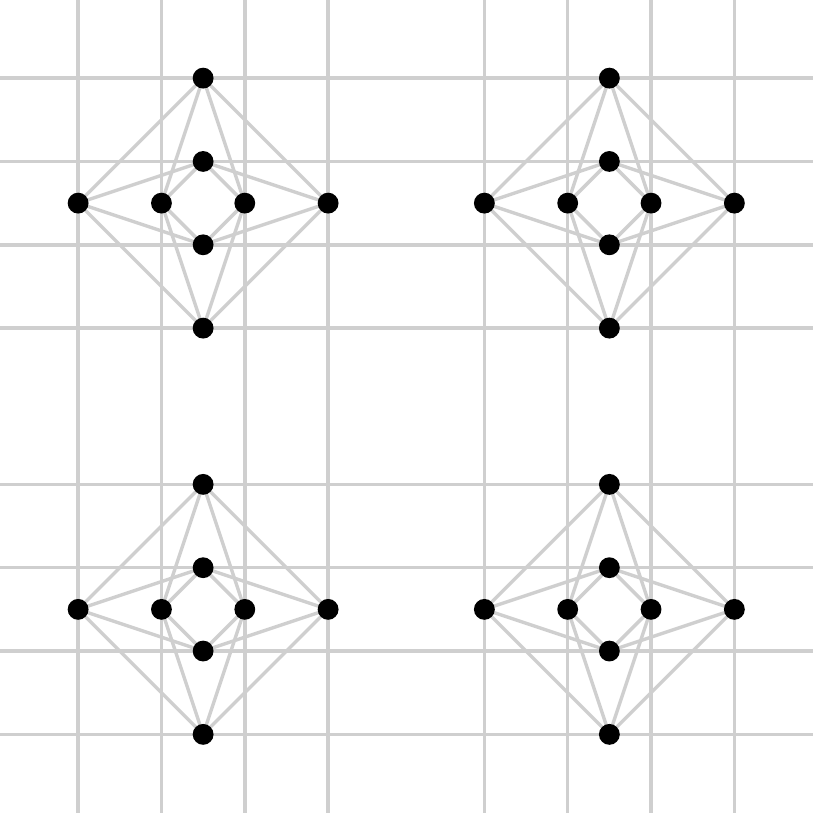}}\subfloat[Triangle]{

\includegraphics[width=0.33\textwidth]{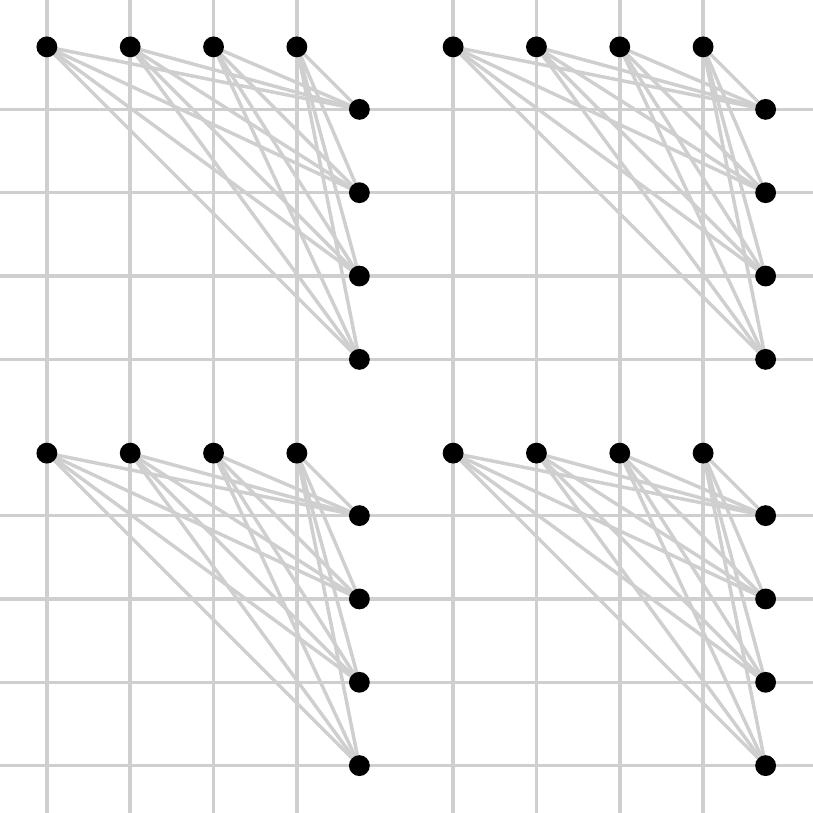}}
\end{figure*}

The 128 qubits of the first commercial quantum annealer (D-Wave One,
released in 2011) were connected by a graph called Chimera (first
defined publicly in 2009 \citep{Neven2009}), which is rather easy
to describe: A 2D array of $K_{4,4}$ graphs, with one `side' of each
$K_{4,4}$ being connected to the same corresponding side on the $K_{4,4}$
cells directly above and below it, and the other side being connected
to the same corresponding side on the $K_{4,4}$ cells to the right
and left of it (see Figure \ref{fig:chimera}). The qubits can couple
to up to 6 other qubits, since each qubit couples to 4 qubits within
its $K_{4,4}$ unit cell, and to 2 qubits in the $K_{4,4}$ cells
above and below it, or to the left and right of it. All commercial
quantum annealers built to date follow this graph connectivity, with
just larger and larger numbers of $K_{4,4}$ cells (see Table 1).

\begin{table}[H]
\caption{Chimera graphs in all commercial quantum annealers to date.}

\begin{tabular*}{1\columnwidth}{@{\extracolsep{\fill}}ccc}
\hline 
\noalign{\vskip2mm}
 & Array of $K_{4,4}$ cells & Total \# of qubits\tabularnewline[2mm]
\hline 
\noalign{\vskip2mm}
D-Wave One & $4\times4$ & 128\tabularnewline
D-Wave Two & $8\times8$ & 512\tabularnewline
D-Wave 2X & $12\times12$ & 1152\tabularnewline
D-Wave 2000Q & $16\times16$ & 2048\tabularnewline[2mm]
\hline 
\end{tabular*}
\end{table}

In 2018, D-Wave announced the construction of a (not yet commercial)
quantum annealer with a greater connectivity than Chimera offers,
and a program (NetworkX) which allows users to generate certain Pegasus
graphs. However, no explicit description of the graph connectivity
in Pegasus has been published yet, so we have had to apply the process
of reverse engineering to determine it, and the following section
describes the algorithm we have established for generating Pegasus.

\vspace{-10mm}

\section{Algorithm for generating Pegasus}

\vspace{-5mm}

\begin{figure*}
\vspace{7mm}
\caption{The Pegasus graph.}

\begin{minipage}[t]{0.62\textwidth}%
\subfloat[A patch of size $(X,Y,Z)=(2,2,3)$ cropped out of a Pegasus graph.
Grey edges\protect \\
are part of the three layers of Chimera graphs, while black and blue
edges form the remainder of the Pegasus graph (the latter connecting
different layers).\label{fig:pegasus}]{

\setbox1=\hbox{\includegraphics[height=4cm]{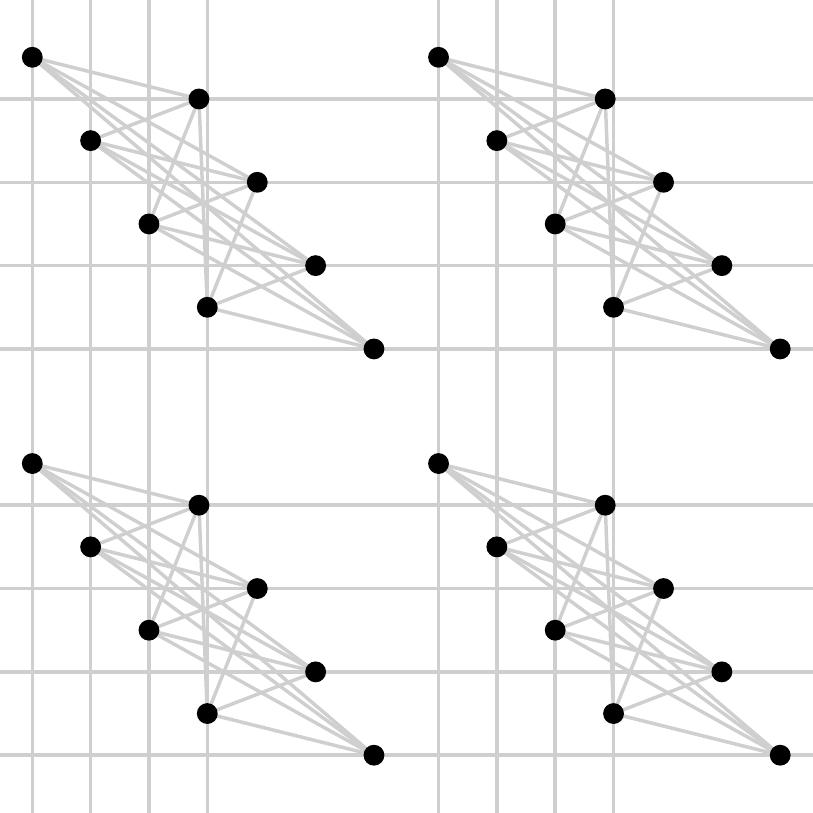}}

\vspace{-6mm}\includegraphics[width=\textwidth]{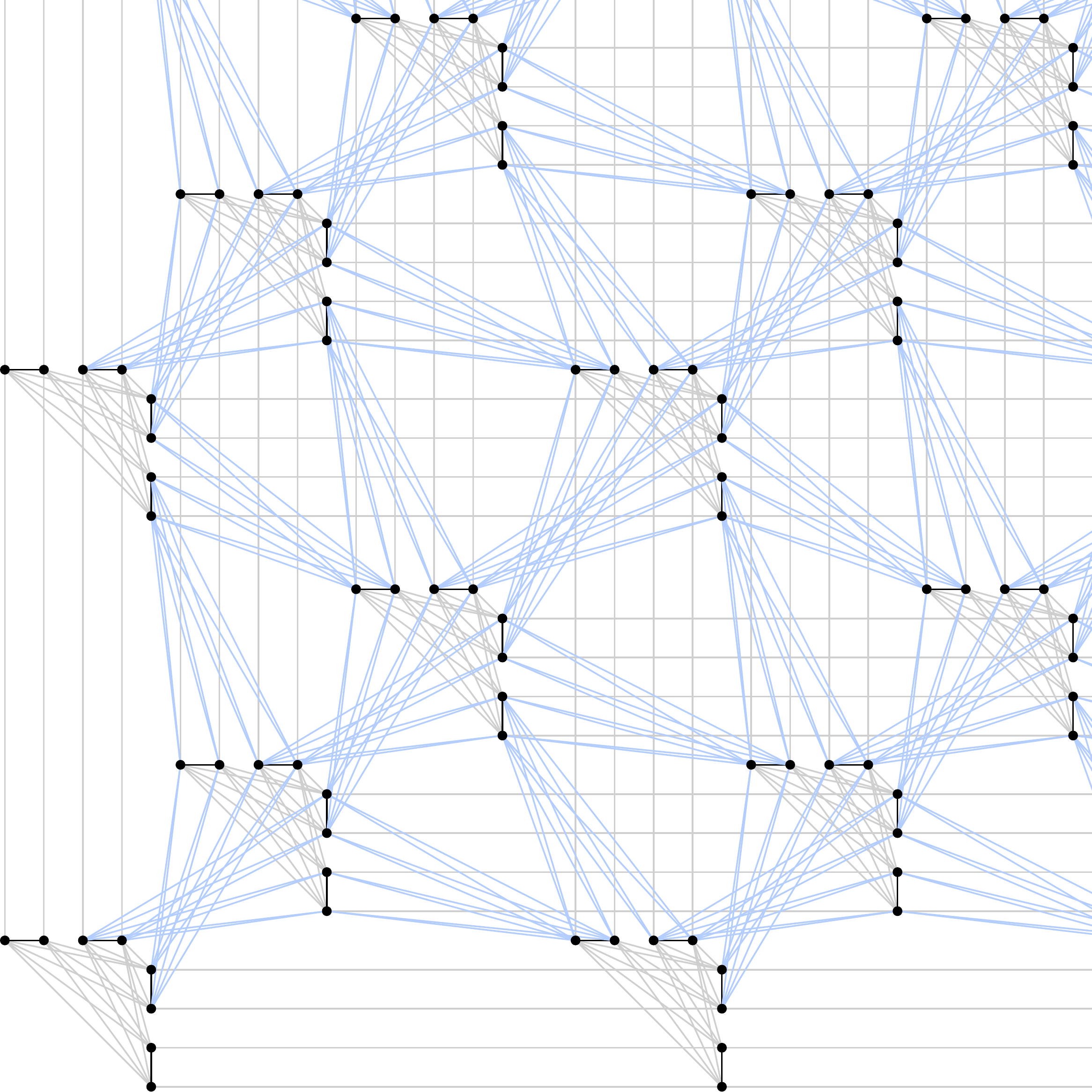}\llap{\hspace{-9.65cm}\fboxsep=0mm\raisebox{7.8cm}{\fcolorbox{gray}{white}{\includegraphics[height=3.3cm]{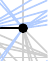}}}\hspace{9.65cm}}

}%
\end{minipage}\hfill{}%
\begin{minipage}[t]{0.355\textwidth}%
\subfloat[Groups of $K_{2,4}$ edges connecting cells belonging to different
Chimera layers, colored according to Eqs. (\ref{eq:z01_j0_i0})-(\ref{eq:z2_j1_i1}).
\label{fig:pegasus64}]{

\vspace{-6mm}

\vspace{-6mm}\includegraphics[width=\textwidth]{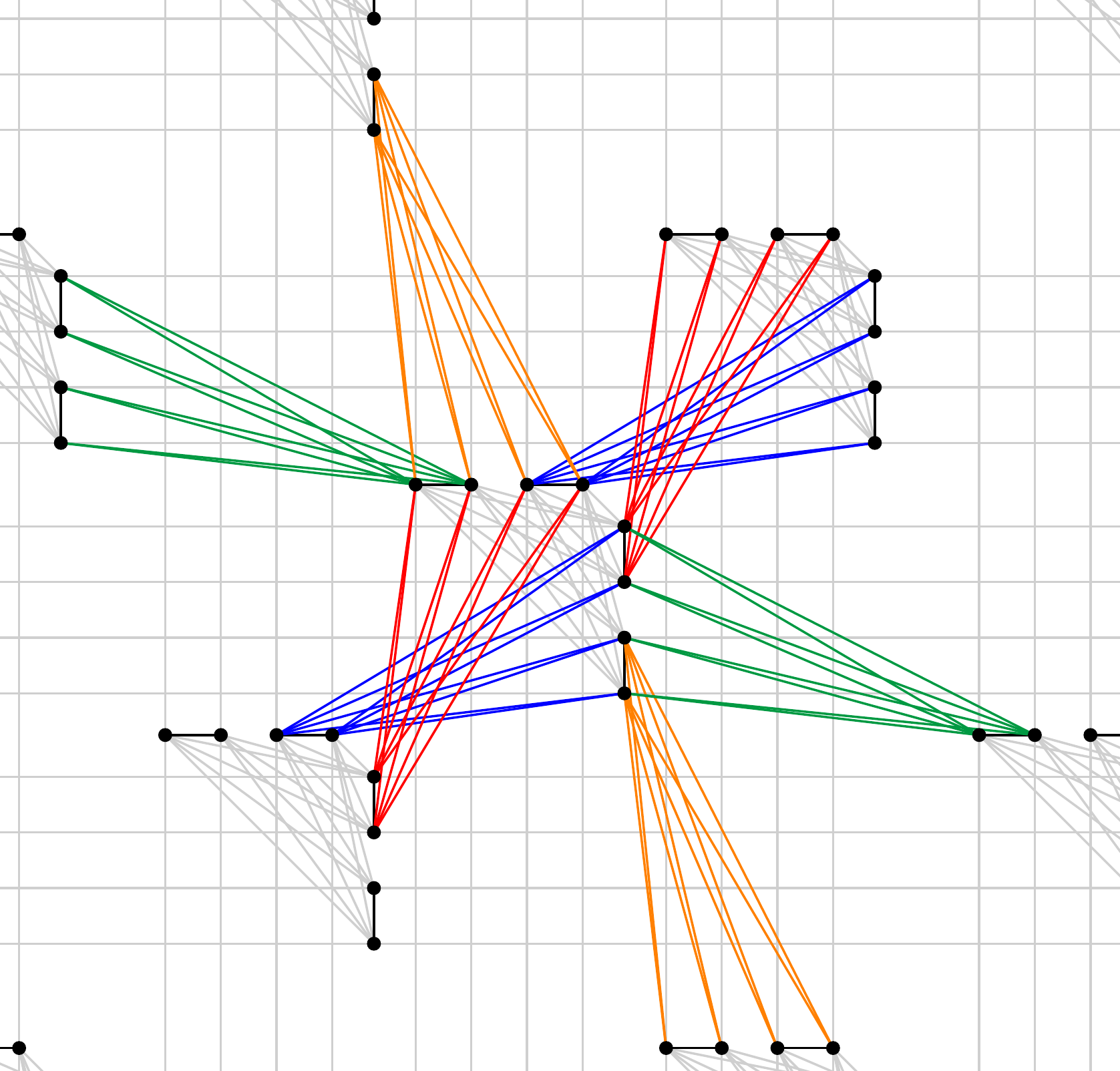}\llap{\hspace{2cm}\fboxsep=0.2mm\raisebox{4.7cm}{\fcolorbox{LightGray}{white}{\includegraphics[height=1.5cm]{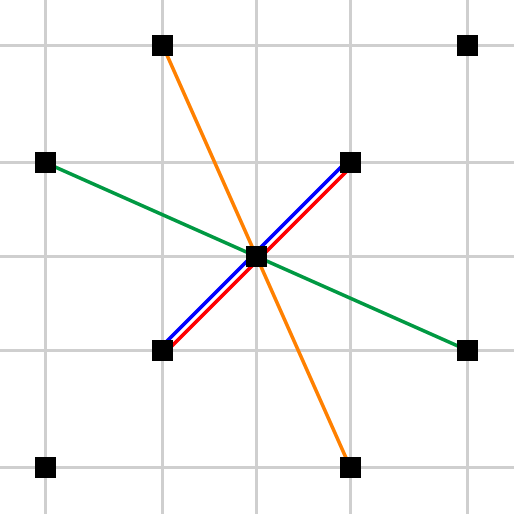}}}\hspace{-0.2cm}}}

\vspace{-2mm}

\subfloat[`Compressed' version of the Pegasus graph, with each cell represented
by one vertex and each $K_{2,4}$ (8 edges) represented by one edge.
\label{fig:pegasusdot}]{

\includegraphics[width=1\textwidth,height=0.55\textwidth]{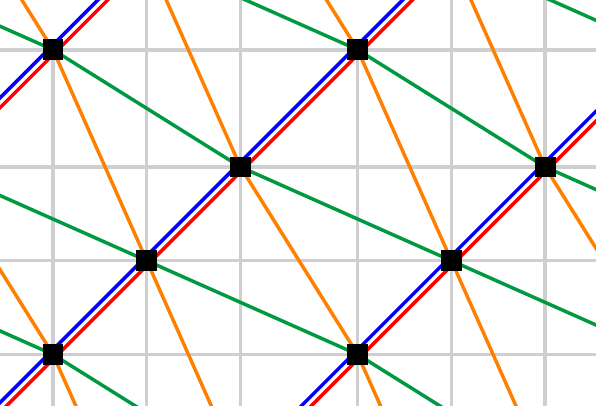}}%
\end{minipage}
\end{figure*}

\subsection{The vertices (qubits)}

Start with $Z$ layers of Chimera graphs, each being an $X\times Y$
array of $K_{4,4}$ cells (therefore we have an $X\times Y\times Z$
array of $K_{4,4}$ cells). The indices $(x,y,z)$ will be used to
describe the location of each cell along the indices corresponding
to the dimension picked from ($X,Y,Z)$.  The values of $X$ and $Y$
have no restriction, but $Z=3$ in Pegasus. Each $K_{4,4}$ cell has
two `sides', labeled $i\in\{0,1\}$, so that there are 4 qubits (vertices)
for every combination: $(x,y,z,i).$ We will arbitrarily label these
4 qubits using two more labels: $j\in\{0,1\}$ and $k\in\{0,1\}$.
Therefore every qubit is associated with 6 indices: $(x,y,z,i,j,k)$,
with their ranges and descriptions given in Table \ref{tab:IndicesInPegasus}. 

In all of the figures in this publication, the origin will be in the
bottom-left corner, the $x$ index will increase in the right-hand
direction, the $y$ index will increase in the upward direction, and
the $z$ index (which indicates which Chimera layer is being considered)
will increase in the direction going upward and rightward simultaneously
(since paper and computer screens are still two-dimensional), see
Figure \ref{fig:pegasus}. Then $i=0$ will represent the left side
in the classic $K_{4,4}$ depiction, or the horizontal vertices in
the diamond-shaped or triangle-shaped depictions; while $i=1$ will
represent the right side in the classic and vertical vertices in the
diamond and triangle.

\vspace{-2mm}

\begin{table}[H]
\caption{Indices used to describe each qubit (vertex) in Pegasus.\label{tab:IndicesInPegasus}}
\vspace{-2mm}
\vspace{-1mm}

\begin{tabular}{ccc}
\hline 
\noalign{\vskip2mm}
Index & Range & Description\tabularnewline[2mm]
\hline 
\noalign{\vskip2mm}
$x$ & $0$ to $X-1$ & Row within a Chimera layer\tabularnewline
$y$ & $0$ to $Y-1$ & Column within a Chimera layer\tabularnewline
$z$ & $0$ to $2$ & Chimera layer\tabularnewline
$i$ & $0,1$ & `Side' within $K_{4,4}$\tabularnewline
$j$ & $0,1$ & First index within each `side' of $K_{4,4}$\tabularnewline
$k$ & $0,1$ & Second index within each `side' of $K_{4,4}$\tabularnewline[2mm]
\hline 
\end{tabular}
\end{table}

\vspace{-2mm}

\vspace{-2mm}

\vspace{-2mm}

\vspace{-2mm}

\vspace{-2mm}

\subsection{The edges (couplings) in Chimera}

\vspace{-2mm}

\subsubsection{Edges forming each $K_{4,4}$ cell}

\vspace{-2mm}

The $K_{4,4}$ cells are given by:
\begin{equation}
(x,y,z,0,j,k)\longleftrightarrow(x,y,z,1,j',k').\label{eq:k44}
\end{equation}
This means for each $K_{4,4}$ cell, all vertices $(j,k)$ for side
$i=0$ are coupled to all vertices $(j^{\prime},k^{\prime})$ of side
$i=1$. \textbf{\emph{For this entire publication, $j^{\prime}$ can
be equal to or different from $j$ (and likewise for $k^{\prime}$and
$k$)}}\textbf{.}

\vspace{-2mm}

\vspace{-2mm}

\vspace{-2mm}

\subsubsection{Edges connecting different $K_{4,4}$ cells}

\vspace{-2mm}

The horizontal lines between $K_{4,4}$ cells in Figure \ref{fig:chimera}
can be described by adding 1 to $x$ while keeping all other variables
constant and setting $i=1$:

\vspace{-2mm}

\vspace{-2mm}

\begin{equation}
(x,y,z,1,j,k)\longleftrightarrow(x+1,y,z,1,j,k),\label{eq:chimera_horizontal}
\end{equation}
and the vertical lines can be described by adding 1 to $y$ while
keeping all other variables constant and setting $i=0$:

\vspace{-2mm}

\vspace{-2mm}

\vspace{-2mm}

\vspace{-2mm}

\begin{equation}
(x,y,z,0,j,k)\longleftrightarrow(x,y+1,z,0,j,k).\label{eq:chimera_vertical}
\end{equation}
For each $z$, Equations (\ref{eq:k44})-(\ref{eq:chimera_vertical})
define edges connecting vertices labeled by $x,y,i,$ $j$ and $k$.
This completes the definition of a Chimera graph. In all figures,
these Chimera edges are grey.

\vspace{-2mm}

\subsection{The \emph{new} edges (couplings) in Pegasus}

\vspace{-2mm}

\subsubsection{New edges added to each $K_{4,4}$ cell:}

\vspace{-2mm}

Pegasus first adds connections within each $K_{4,4}$ cell, given
by simply coupling each vertex labeled as $k=0$ to its $k=1$ counterpart
with all other variables unchanged:
\begin{equation}
(x,y,z,i,j,0)\longleftrightarrow(x,y,z,i,j,1).\label{eq:uc_add}
\end{equation}
These edges are drawn black in the figures.

\subsubsection{Edges connecting different $K_{4,4}$ cells:}

\vspace{-2mm}

\vspace{-2mm}

The rest of the new connections in Pegasus come from connecting the
$K_{4,4}$ cells between different layers (different $z$) of Chimera
graphs. The qubits of a $K_{4,4}$ cell located at coordinates $(x,y,z)$
will be connected to $6$ different $K_{4,4}$ cells on the other
Chimera layers, with 64 edges in the form of 8 different $K_{2,4}$
graphs: 1 $K_{2,4}$ graph (8 edges) for each of 4 different connecting
$K_{4,4}$ cells, and 2 $K_{2,4}$ graphs (16 edges) for the other
2 connecting $K_{4,4}$ cells. 

Figure \ref{fig:pegasus64} shows these 64 edges, and the 8 groups
of 8-edge $K_{2,4}$ graphs connecting the central cell to 6 others
are shown with 4 different colors. This is because we have found a
set of 4 convenient rules that can be repeated to generate the entire
Pegasus graph, and it is found that 4 of the $K_{2,4}$ graphs can
be generated by applying these 4 rules to the central cell, and the
other 4 of the $K_{2,4}$ graphs can be generated by applying these
4 rules to other $K_{4,4}$ cells (or by applying the 4 rules to the
central cell again, but in reverse). We will now describe the rules.

\bigskip{}

First, all edges are between vertices of one $K_{4,4}$ side $i$
and its complementary side $\bar{i}$ in a different $K_{4,4}$ (so
$i=0$ vertices are coupled to $i=1$ vertices of a $K_{4,4}$ in
a different layer). In fact all connections will be of the form $(i,j,k)\longleftrightarrow(\bar{i},j^{\prime},k^{\prime})$
where $j^{\prime}$ and $k^{\prime}$ can be any value in $\{0,1\}$.
\textbf{\emph{All edges can actually be described using just a one-line
rule:}}

\vspace{-4mm}

\begin{multline*}
\,\,\,\,\,\,\,\,\,\,\,\,\,\,\,\,\,\,\,\,\,\,\,\,\,\,\,\,\,\,\,\,\,\,\,\,\,\,\,\,\,\,\,\,\,\,\,\,\left(x,y,z,i,j,k\right)
\end{multline*}

\vspace{-8mm}

\begin{equation}
\longleftrightarrow\label{eq:motherOfAllEdgeRelations}
\end{equation}

\vspace{-7mm}
\[
\left(x-j\bar{i}+\delta_{z2},y-ji+\delta_{z2},(z+1)\negthinspace\negthinspace\negthinspace\negthinspace\negthinspace\mod\negthinspace3,\bar{i},j',k'\right).
\]

~~\rule[0.5ex]{0.9\columnwidth}{1pt}

\smallskip{}

For the layers labeled $z=0$ or 1 (meaning that $\delta_{z2}=0)$,
we have:

\vspace{-5mm}

\begin{equation}
(x,y,z,i,j,k)\longleftrightarrow(x-j\bar{i},y-ji,z+1,\bar{i},j',k').\label{eq:gen_z01}
\end{equation}
Substituting $j=0$ into Eq. (\ref{eq:gen_z01}) tells us that all
$j=0$ vertices of a $K_{4,4}$ cell are connected to all vertices
of the opposite side $\bar{i}$ in the $K_{4,4}$ cell with the same
$(x,y)$, but in the next layer $z+1$: 

\vspace{-5mm}

\begin{equation}
(x,y,z)\longleftrightarrow(x,y,z+1),\,\text{for}\,\text{\ensuremath{z\in\{0,1\},j=0.}}
\end{equation}

\noindent Substituting $j=1$ into Eq. (\ref{eq:gen_z01}) tells us
that all $j=1$ vertices are also connected to all $j'$ and $k'$
vertices in the opposite side  $\bar{i}$ in a $K_{4,4}$ cell in
the next layer $z+1$, but with $x$ and $y$ coordinates shifted
by 1 in the following way: 

\vspace{-5mm}

\begin{align}
\negthinspace\negthinspace(x,y,z) & \longleftrightarrow(x-\bar{i},y-i,z+1)\,\text{for}\,\text{\ensuremath{z\in\{0,1\},j=1.}}
\end{align}

~~\rule[0.5ex]{0.9\columnwidth}{1pt}

\smallskip{}

For the $z=2$ layer ($\delta_{z2}=1)$, we have from Eq. (\ref{eq:motherOfAllEdgeRelations}):
\begin{equation}
(x,y,2,i,j,k)\longleftrightarrow(x-j\bar{i}+1,y-ji+1,0,\bar{i},j',k').\label{eq:gen_z2}
\end{equation}
Substituting $j=0$ into Eq. (\ref{eq:gen_z2}) tells us that all
$j=0$ vertices of a $K_{4,4}$ cell are connected to the $K_{4,4}$
cells in the $z=0$ layer, but with $x$ and $y$ coordinates \emph{both}
shifted by 1:

\vspace{-2mm}

\vspace{-2mm}

\vspace{-2mm}

\begin{align}
(x,y,2) & \longleftrightarrow(x+1,y+1,0),\,\text{for}\text{\,\ensuremath{j=0.}}\label{eq:}
\end{align}
Substituting $j=1$ into Eq. (\ref{eq:gen_z2}) tells us that all
$j=1$ vertices are connected in the following way:

\vspace{-2mm}

\vspace{-2mm}

\begin{align}
(x,y,2) & \longleftrightarrow(x+i,y+\bar{i},0),\,\text{for\,\ensuremath{j=1}}.
\end{align}

\vspace{-2mm}

\rule[0.5ex]{0.9\columnwidth}{1pt}

\smallskip{}

We now explicitly write down the 4 rules which lead to the grouping
scheme depicted in Figure \ref{fig:pegasus64} (these rules are different
depending on whether $z\in\{0,1\}$ or $z=2$, so there are actually
8 relations):\vspace{-2mm}

\vspace{-2mm}

\vspace{-2mm}

\begin{align}
\negthinspace\negthinspace\negthinspace\negthinspace\negthinspace\negthinspace\textcolor{blue}{(x,y,z,0,0,k)}\, & \textcolor{blue}{\leftrightarrow}\,\textcolor{blue}{(x,y,z+1,1,j',k'),\,\,\,\,\,\,\,\,\,\,\,z\in\{0,1\}}\negthinspace\negthinspace\label{eq:z01_j0_i0}\\
\textcolor{red}{\negthinspace\negthinspace\negthinspace\negthinspace\negthinspace\negthinspace(x,y,z,1,0,k)}\, & \textcolor{red}{\leftrightarrow}\textcolor{red}{\,(x,y,z+1,0,j',k'),\,\,\,\,\,\,\,\,\,\,\,z\in\{0,1\}\negthinspace}\negthinspace\label{eq:z01_j0_i1}\\
\textcolor{Green}{\negthinspace\negthinspace\negthinspace\negthinspace\negthinspace\negthinspace(x,y,z,0,1,k)}\, & \textcolor{Green}{\leftrightarrow}\textcolor{Green}{\,(x-1,y,z+1,1,j',k'),\,z\in\{0,1\}}\negthinspace\negthinspace\label{eq:z01_j1_i0}\\
\textcolor{Orange}{\negthinspace\negthinspace\negthinspace\negthinspace\negthinspace\negthinspace(x,y,z,1,1,k)}\, & \textcolor{Orange}{\leftrightarrow}\textcolor{Orange}{\,(x,y-1,z+1,0,j',k'),\,z\in\{0,1\}}\negthinspace\negthinspace\label{eq:z01_j1_i1}\\
\nonumber \\
\negthinspace\negthinspace\negthinspace\negthinspace\negthinspace\negthinspace\textcolor{blue}{(x,y,2,0,0,k)}\, & \textcolor{blue}{\leftrightarrow}\textcolor{blue}{\,(x+1,y+1,0,1,j',k'),}\negthinspace\negthinspace\label{eq:z2_j0_i0}\\
\textcolor{red}{\negthinspace\negthinspace\negthinspace\negthinspace\negthinspace\negthinspace(x,y,2,1,0,k)}\, & \textcolor{red}{\leftrightarrow}\textcolor{red}{\,(x+1,y+1,0,0,j',k'),}\negthinspace\negthinspace\label{eq:z2_j0_i1}\\
\textcolor{Green}{\negthinspace\negthinspace\negthinspace\negthinspace\negthinspace\negthinspace(x,y,2,0,1,k)}\, & \textcolor{Green}{\leftrightarrow}\,\textcolor{Green}{(x,y+1,0,1,j',k'),}\negthinspace\negthinspace\label{eq:z2_j1_i0}\\
\textcolor{Orange}{\negthinspace\negthinspace\negthinspace\negthinspace\negthinspace\negthinspace(x,y,2,1,1,k)}\, & \textcolor{Orange}{\leftrightarrow\,}\textcolor{Orange}{(x+1,y,0,0,j',k').}\negthinspace\negthinspace\label{eq:z2_j1_i1}
\end{align}

\vspace{-2mm}
\vspace{-2mm}

\vspace{-2mm}

\section{Comparison to Chimera}

\vspace{-2mm}

\vspace{-2mm}

\subsection{Degree of the vertices}

\vspace{-2mm}

If we look for example at the cell at position $(x,y,z)=(1,1,1)$
in Fig. \ref{fig:pegasus}, we can find that vertices have a degree
of \textbf{\emph{fifteen: }}The 6 grey edges that would regularly
be in Chimera (4 to form the $K_{4,4}$ cell and 2 to connect to $K_{4,4}$
cells above/left and below/right), then there is 1 Pegasus edge added
\emph{within} one $K_{4,4}$ cell according to Eq. (\ref{eq:uc_add}),
then we have 8 more Pegasus edges for connecting $K_{4,4}$ cells
of different Chimera layers $z$, from Eqs. (\ref{eq:z01_j0_i0})-(\ref{eq:z01_j1_i1})
(or from Eqs. (\ref{eq:z2_j0_i0})-(\ref{eq:z2_j1_i1}) if we were
starting with a cell in the $z=2$ layer). Therefore the degree (which
is 15) has increased by a factor of 2.5 when compared to the degree
of Chimera (which is 6).

In Fig. \ref{fig:pegasus}, the cells at the boundary, such as at
position $(x,y,z)=(0,0,0)$, do not show a degree of 15, just as the
cells of Chimera at the boundary would not show a degree of 6.

\vspace{-2mm}

\vspace{-2mm}

\vspace{-2mm}

\subsection{Non-planarity}

\vspace{-2mm}

We note that certain binary optimization problems forming planar graphs
can be solved on a classical computer with a number of operations
that scales polynomially with the number of binary variables, with
the blossom algorithm \cite{Edmonds1965,*Edmonds1965a}. Therefore
it is important that the qubits of a quantum annealer are connected
by a non-planar graph. The $K_{4,4}$ cells of Chimera are already
sufficient to make all commercial D-Wave annealers non-planar. However,
if each $K_{4,4}$ cell of a Chimera's physical qubits were to encode
just one logical qubit (in for example, an extreme case of minor embedding),
then Chimera would be planar. While all added edges in Pegasus that
connect different Chimera layers are of the form $K_{2,4}$, which
itself is planar; these $K_{2,4}$ edges connect cells of different
planes of chimeras in a non-planar way, such that even if each cell
were to represent one logical qubit, \emph{these logical qubits would
still form a non-planar graph in Pegasus.} This should expand the
number of binary optimization problems that cannot yet be solved in
polynomially time, that can potentially be embedded onto a D-Wave
annealer. 

\vspace{-2mm}
\vspace{-2mm}
\vspace{-2mm}

\vspace{-0.5mm}

\subsection{Embedding}

\vspace{-0.5mm}

\vspace{-2mm}

We have written an entire paper on the minor-embedding of quadratization
\citep{Dattani2019} gadgets onto Chimera and Pegasus \citep{Dattani2019a}.
One highlight of that work is the fact that \emph{all }quadratization
gadgets for single cubic terms which require one auxiliary qubit,
can be embedded onto Pegasus with no further auxiliary qubits because
Pegasus contains $K_{4},$ which means that all three logical qubits
and the auxiliary qubit can be connected in any way, without any minor-embedding.
We refer the reader to that paper \citep{Dattani2019a} for more thorough
details about the advantage of Pegasus over Chimera for the minor-embedding
of quadratization gadgets. 

\vspace{-2mm}

\vspace{-2mm}

\section{Open source code for generation of Pegasus figures}

All figures in this publication and its supplemental material can
be generated (in vector graphic form) using our open source and customizable
code which should be cited as Ref. \citep{Szalay2018}. The user can
choose which type of Pegasus graph; the dimensions $X$ and $Y$;
the edge colors and widths, the vertex colors and widths; among other
things (see the manual to Ref. \citep{Szalay2018} for details).

\vspace{-2mm}

\vspace{-2mm}

\section*{Acknowledgments}

\vspace{-2mm}

We gratefully thank Kelly Boothby of D-Wave for her time spent in
verifying the correctness of our algorithm. NC was funded by EPSRC
(Project: EP/S00114X); SzSz by the NRDIF (NKFIH-K120569 and Quantum
Technology National Excellence Program 2017-1.2.1-NKP-2017-00001)
and the HAS (János Bolyai Research Scholarship and ``Lendület'' Program).

\vspace{-2mm}
\foreignlanguage{british}{}\vspace{-2mm}
\vspace{-1mm}

\selectlanguage{british}%
{\scriptsize{}\bibliographystyle{apsrev4-1}
\bibliography{\string"/home/nike/pCloud Sync/library\string"}
}{\scriptsize\par}

\selectlanguage{english}%
\vspace{-2mm}

\clearpage{}

\selectlanguage{british}%

\part*{\foreignlanguage{english}{}}

\selectlanguage{english}%
\newpage{}

\newgeometry{top=15mm,bottom=0mm,left=8mm,right=8mm}
\begin{center}
\part*{\onecolumngrid \centering \vspace{-15mm} Supplementary Material}

\centering\captionsetup[subfigure]{justification=centering}
\begin{center}
\vspace{-5mm}
\par\end{center}

\begin{center}
\begin{figure}[H]
\caption{\centering$K_{4,4}$ cells (Chimera).}

\vspace{1mm}

\subfloat[Tilted classic\vspace{2mm}
]{\centering

~~~~~~~~~~~~~~~~\includegraphics[width=0.22\textwidth]{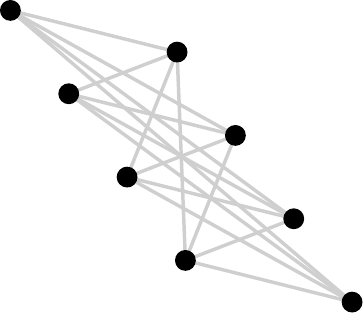}}~~~~~~~~\subfloat[Diamond\vspace{2mm}
]{

\includegraphics[width=0.22\textwidth]{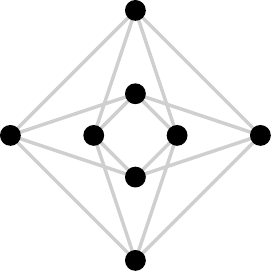}}~~~~~~~~\subfloat[Triangle\vspace{2mm}
]{

\includegraphics[width=0.22\textwidth]{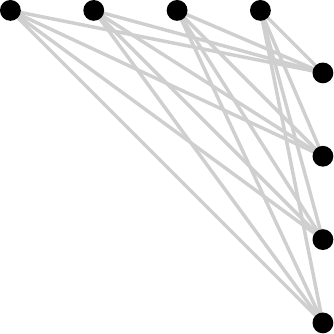}}
\end{figure}
\par\end{center}

\begin{center}
\vspace{-20mm}
\par\end{center}

\begin{center}
\begin{figure}[H]
\caption{\centering$2\times2$ arrays of $K_{4,4}$ cells in Chimera formation.}

\subfloat[Tilted classic\vspace{2mm}
]{

\includegraphics[width=0.24\textwidth]{metapostfigs/fig_ChimeraStdRect5x5_crop}}~~\subfloat[Diamond\vspace{2mm}
]{

\includegraphics[width=0.24\textwidth]{metapostfigs/fig_ChimeraDmdRect5x5_crop}}~~\subfloat[Triangle\vspace{2mm}
]{

\includegraphics[width=0.24\textwidth]{metapostfigs/fig_ChimeraTrpRect5x5_crop}}~~\subfloat[Compressed\vspace{2mm}
]{

\includegraphics[width=0.24\textwidth]{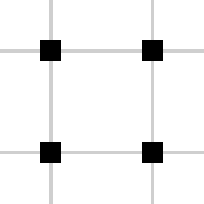}}
\end{figure}
\par\end{center}

\begin{center}
\vspace{-20mm}
\par\end{center}

\begin{center}
\begin{figure}[H]
\caption{\centering$5\times5$ arrays of $K_{4,4}$ cells in Chimera formation.}

\subfloat[Tilted classic\vspace{2mm}
]{

\includegraphics[width=0.24\textwidth]{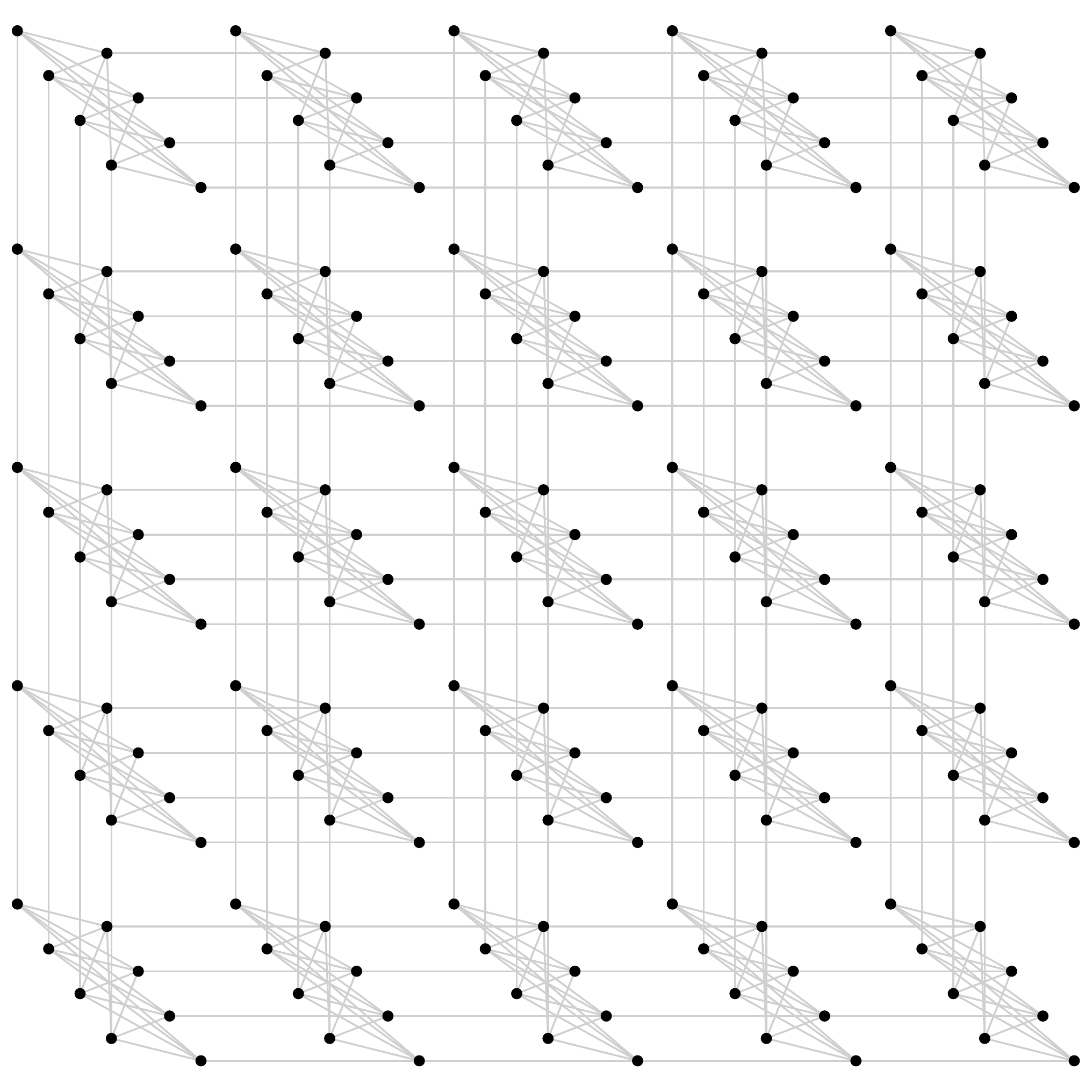}}~~\subfloat[Diamond\vspace{2mm}
]{

\includegraphics[width=0.24\textwidth]{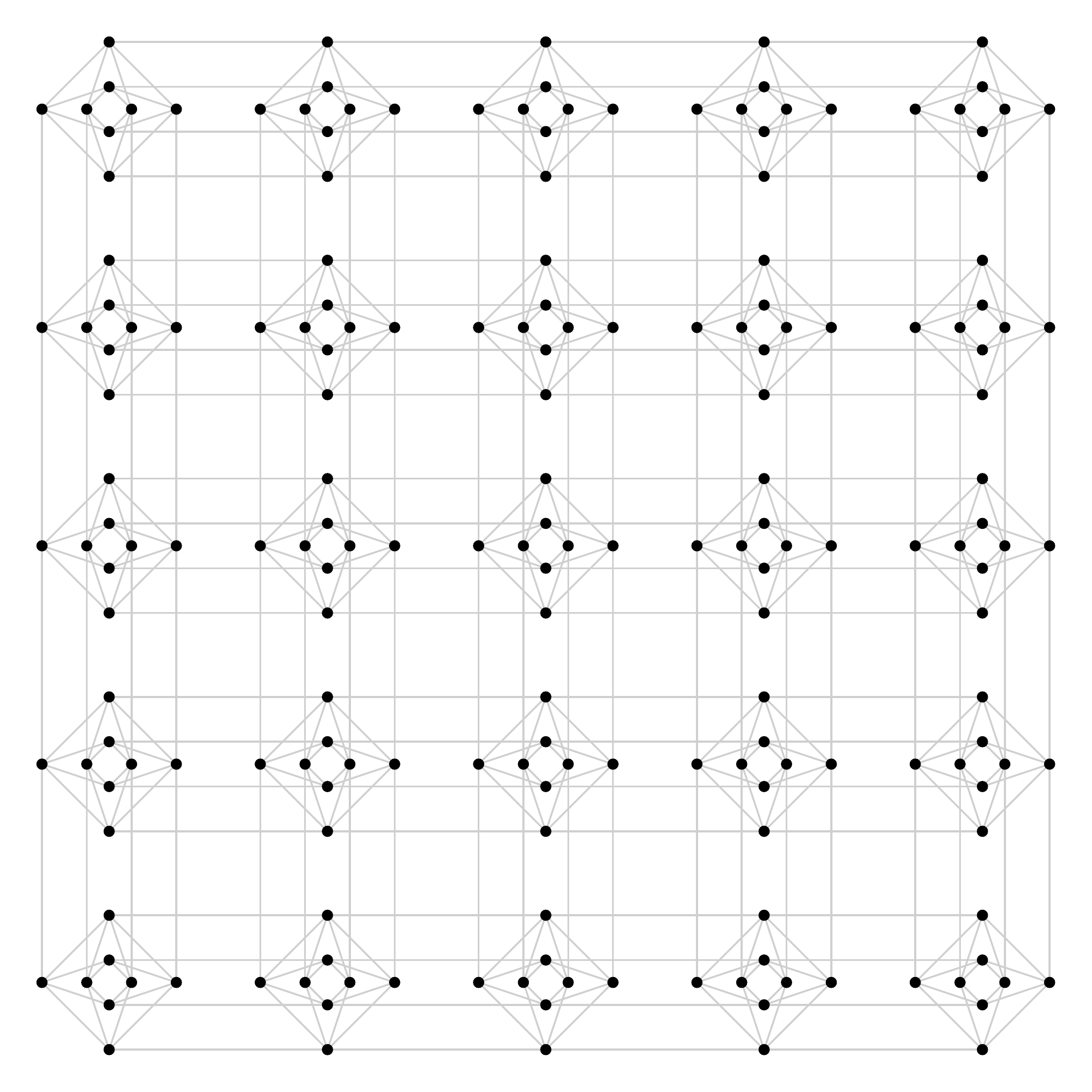}}~~\subfloat[Triangle\vspace{2mm}
]{

\includegraphics[width=0.24\textwidth]{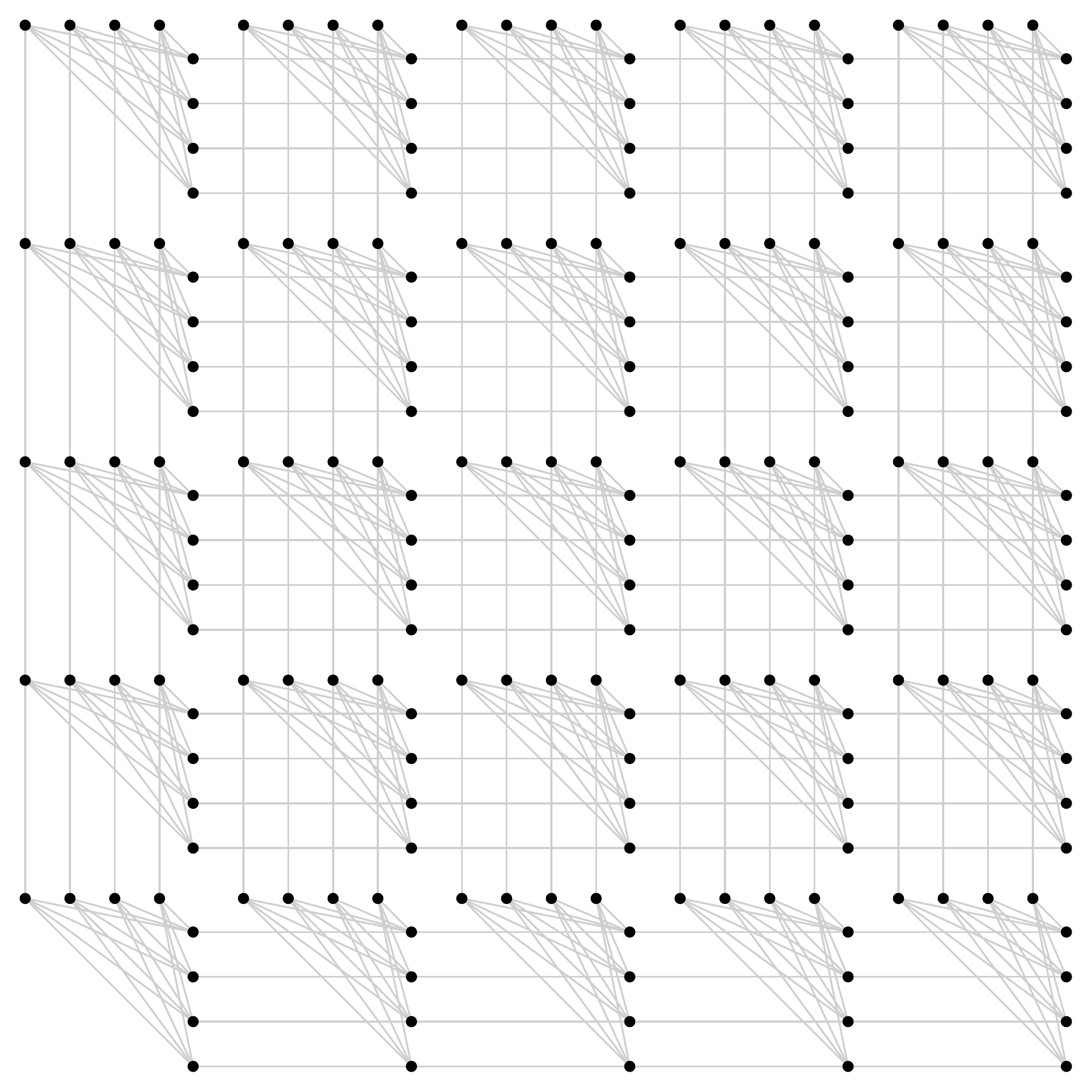}}~~\subfloat[Compressed\vspace{2mm}
]{

\includegraphics[width=0.24\textwidth]{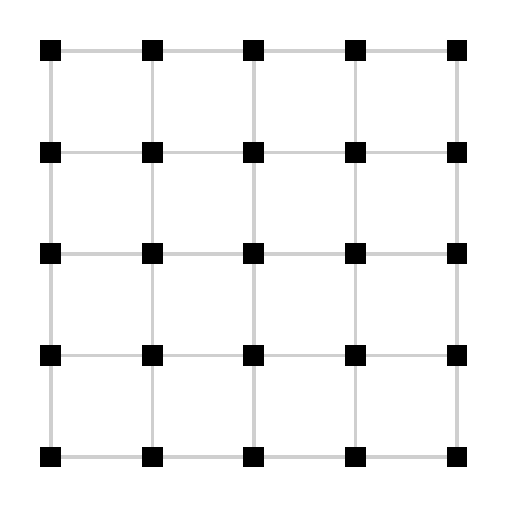}}
\end{figure}
\par\end{center}

\begin{center}
\vspace{-20mm}
\par\end{center}

\begin{center}
\begin{figure}[H]
\caption{\centering$K_{4,4}$ cells (Pegasus).}

\subfloat[Tilted classic\vspace{2mm}
]{

~~~~~~~~~~~~~~~~~\includegraphics[width=0.22\textwidth]{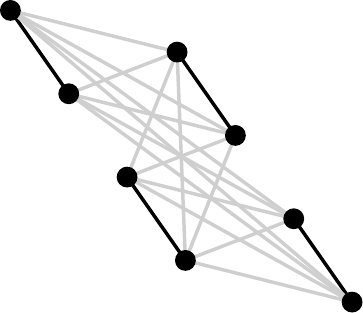}}~~~~~~~~\subfloat[Diamond\vspace{2mm}
]{

\includegraphics[width=0.22\textwidth]{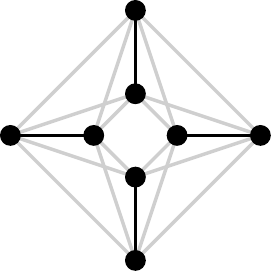}}~~~~~~~~\subfloat[Triangle\vspace{2mm}
]{

\includegraphics[width=0.22\textwidth]{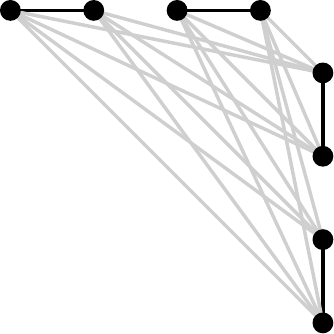}}
\end{figure}
\par\end{center}

\begin{center}
\newpage{}
\par\end{center}

\noindent \begin{center}
\par\end{center}

\begin{center}
\begin{figure}[H]
\vspace{-10mm}
\caption{\centering64 inter-layer edges in color (Pegasus).}

\vspace{15mm}

\subfloat[Tilted classic\vspace{2mm}
]{

\includegraphics[width=0.48\textwidth]{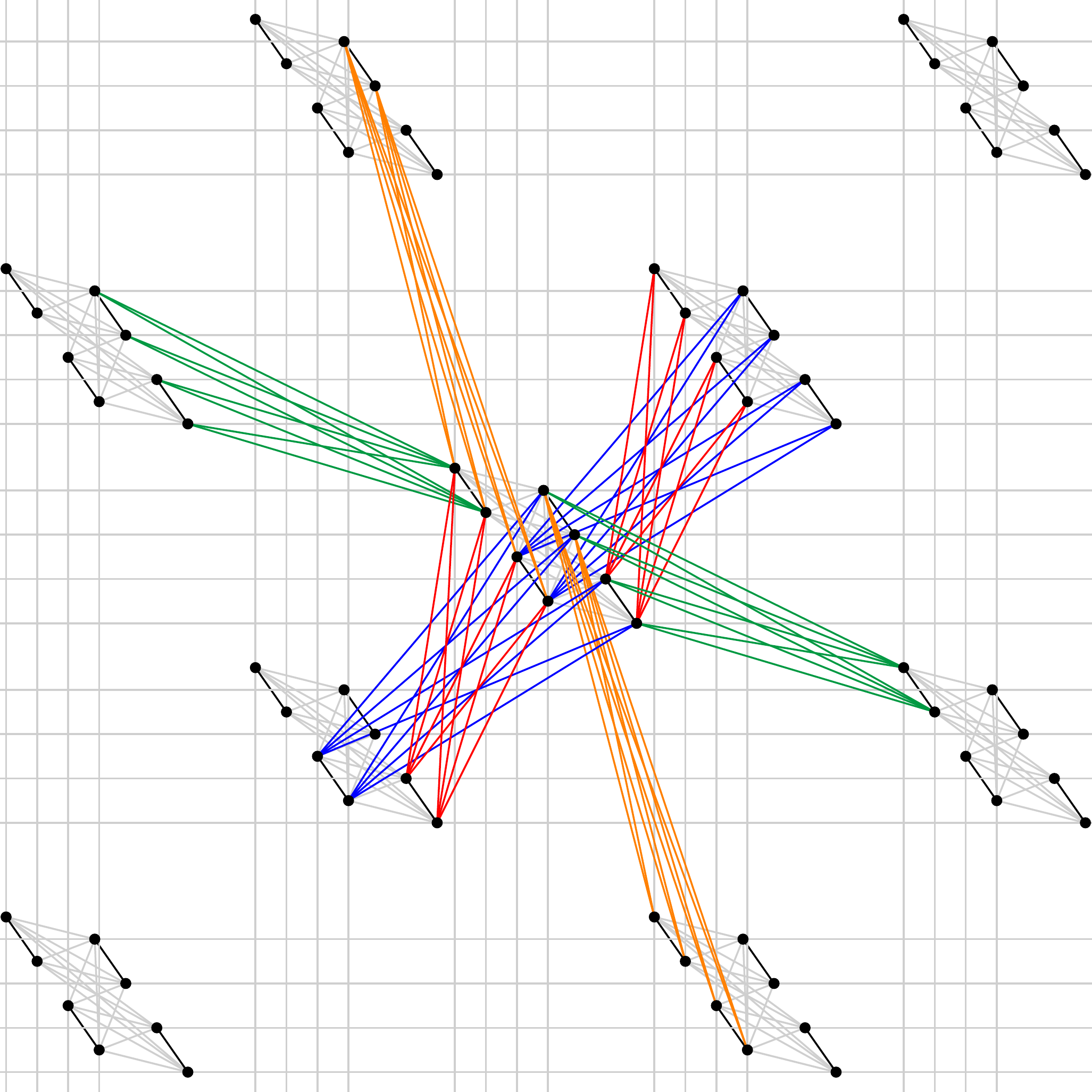}}~~~~~~~~~~\subfloat[Diamond\vspace{2mm}
]{

\includegraphics[width=0.48\textwidth]{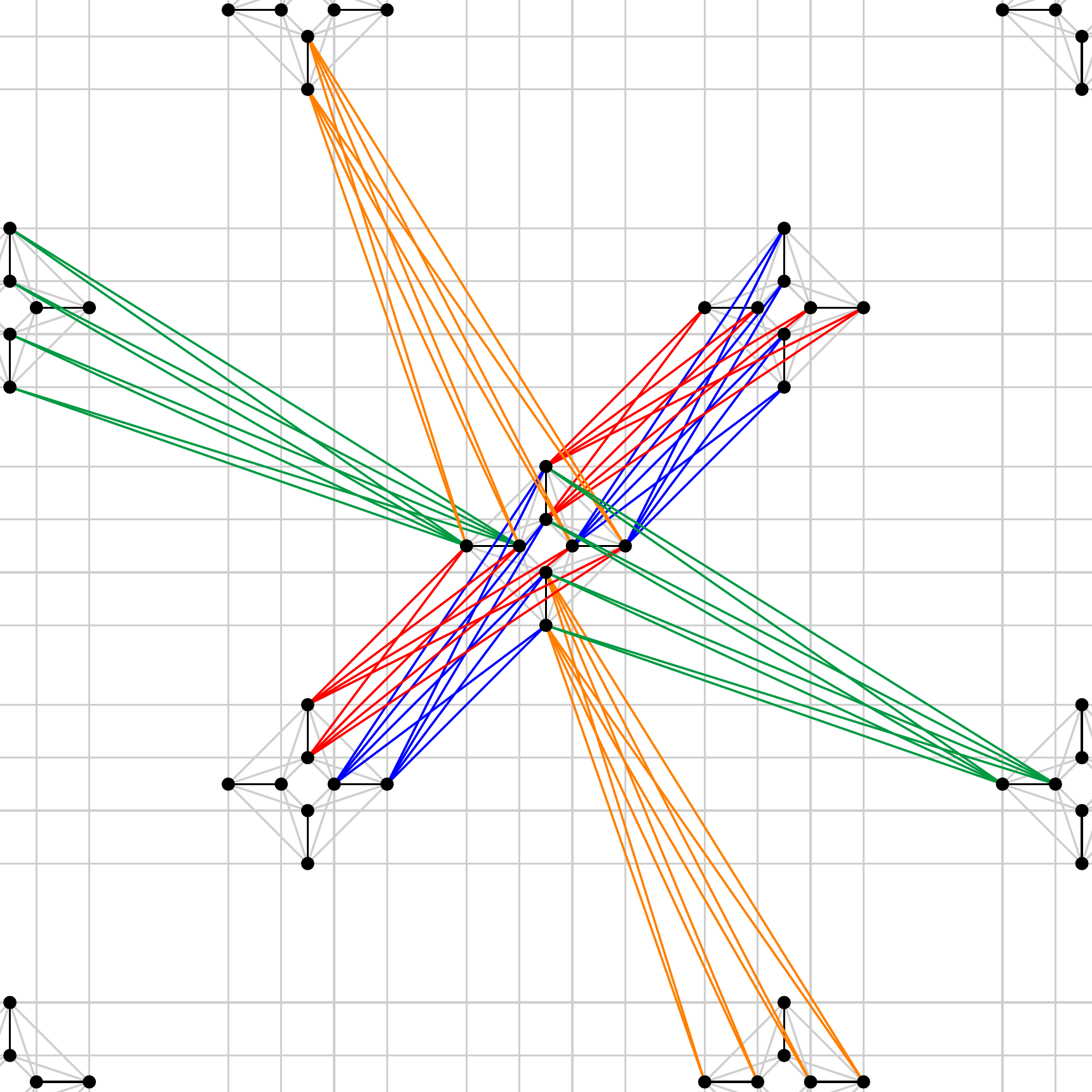}}~~~~~~~~~~

\vspace{10mm}

\subfloat[Triangle\vspace{2mm}
]{

\includegraphics[width=0.48\textwidth]{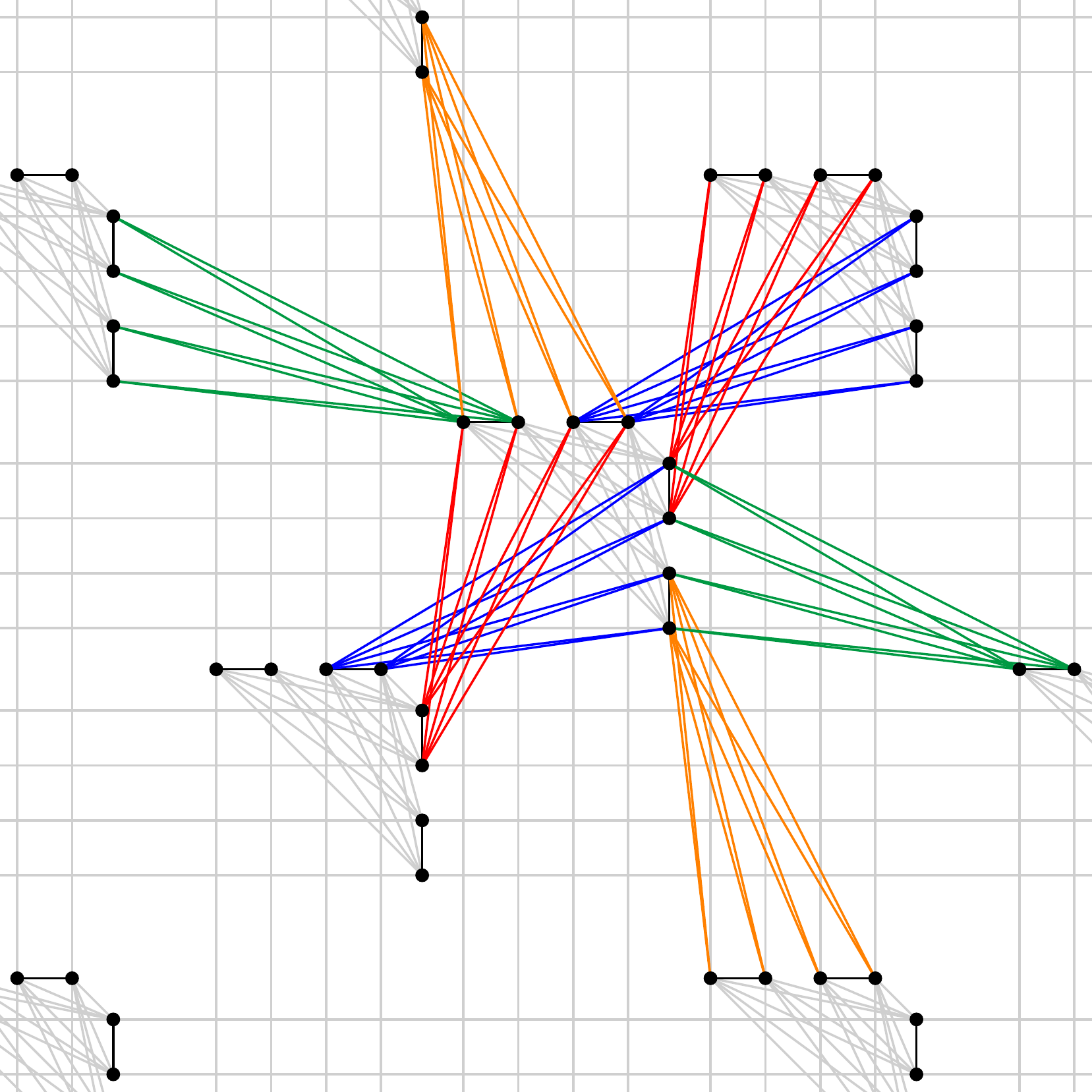}}~~~~~~~~~~\subfloat[Compressed\vspace{2mm}
]{

\includegraphics[width=0.48\textwidth]{metapostfigs/fig_PegasusDotRect5x5one_crop}}
\end{figure}
\par\end{center}

\begin{center}
\newpage{}
\par\end{center}

\begin{center}
\begin{figure}[H]
\vspace{10mm}

\caption{\centering64 inter-layer edges in color within an $(X,Y,Z)=(5,5,3)$
lattice (Pegasus).}

\vspace{15mm}

\subfloat[Tilted classic\vspace{2mm}
]{

\includegraphics[width=0.48\textwidth]{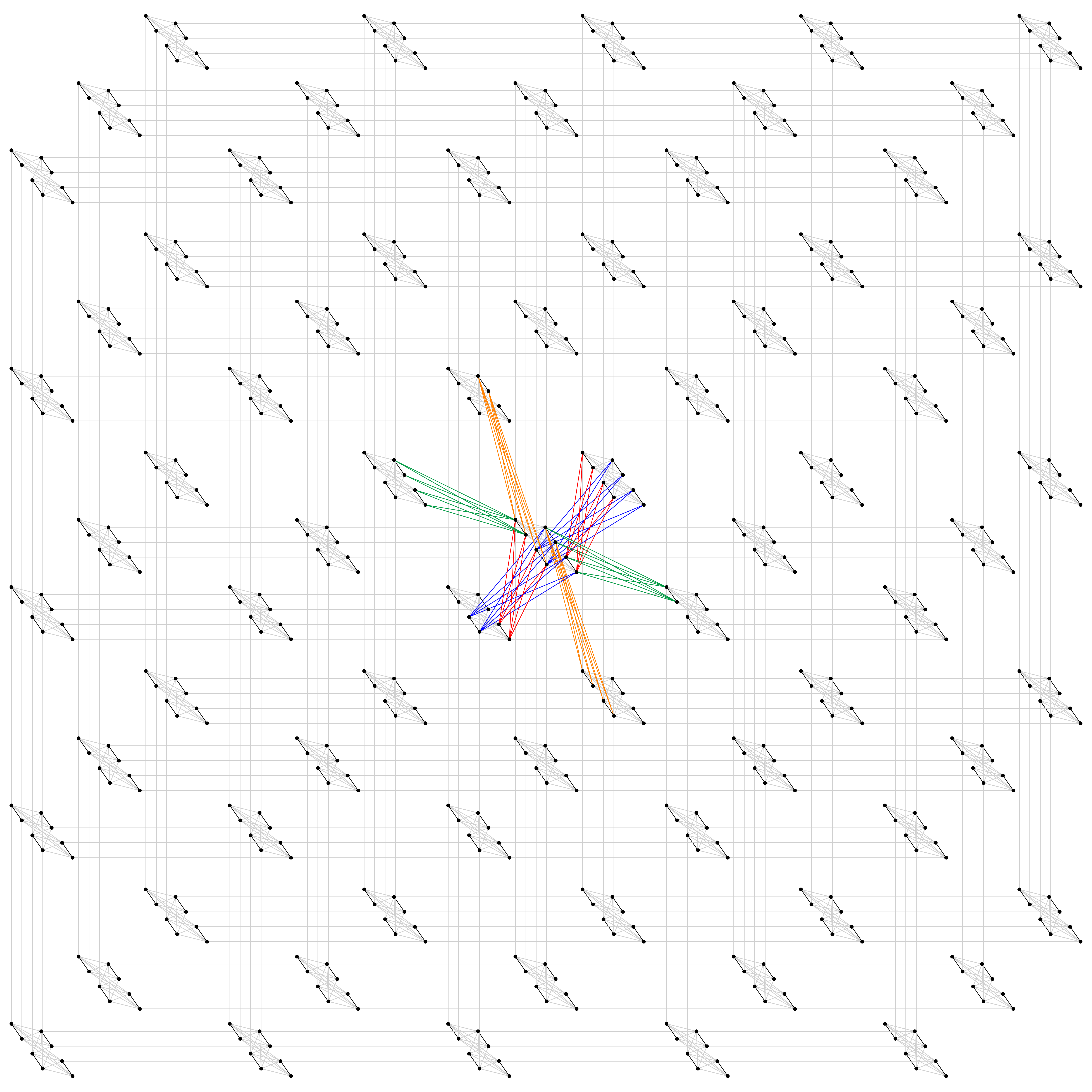}}~~~~~~~~~~\subfloat[Diamond\vspace{2mm}
]{

\includegraphics[width=0.48\textwidth]{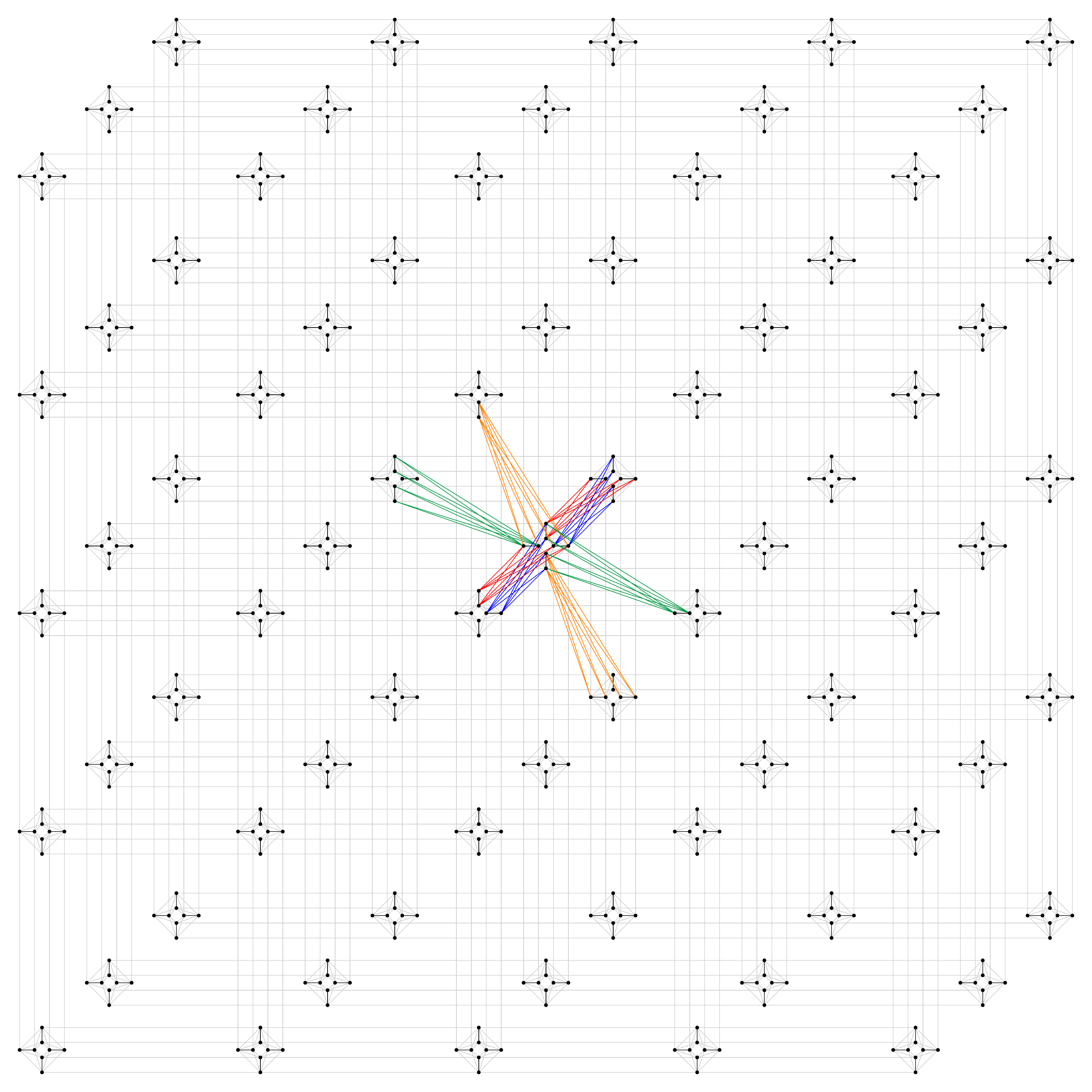}}~~~~~~~~~~

\subfloat[Triangle\vspace{20mm}
]{

\includegraphics[width=0.48\textwidth]{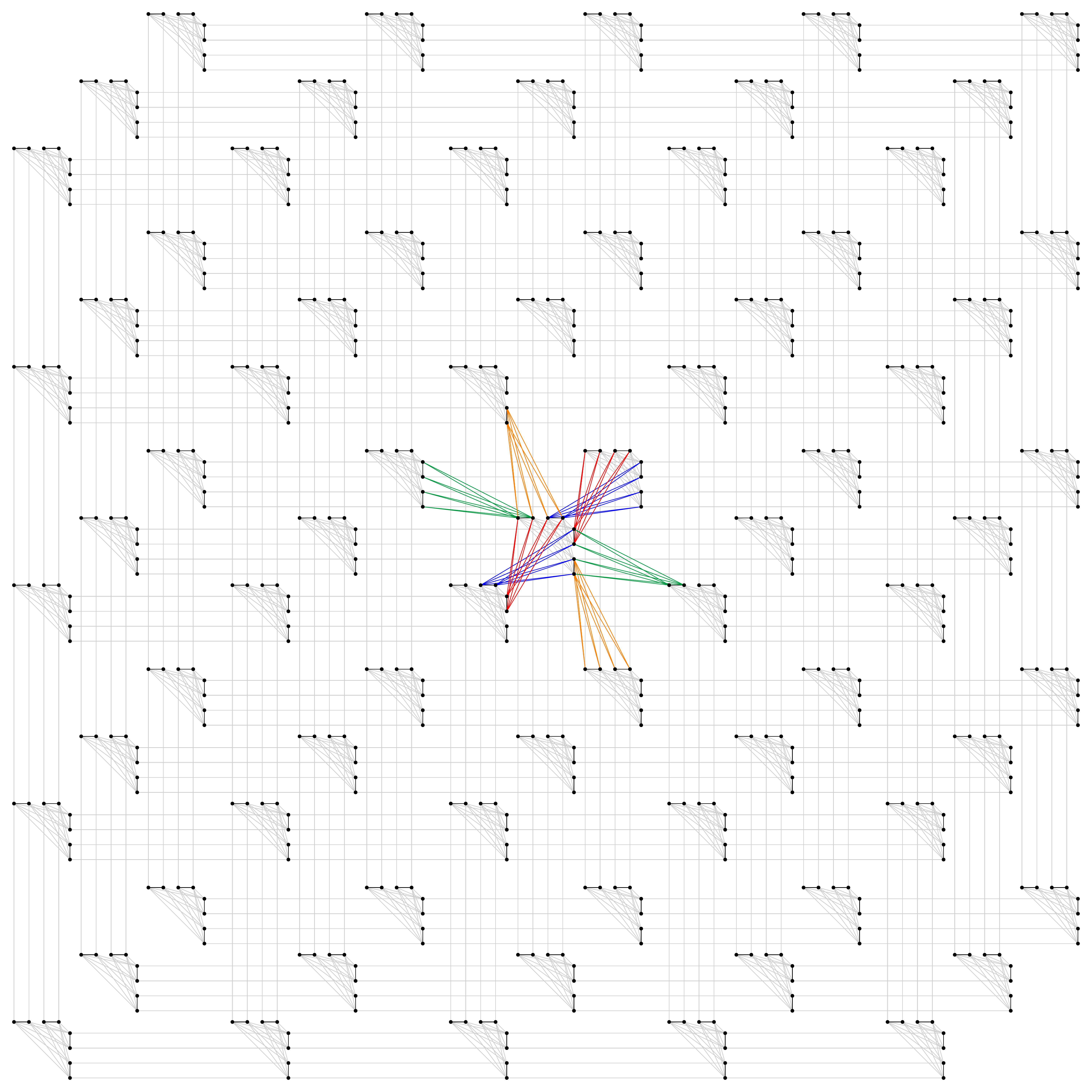}}~~~~~~~~~~\subfloat[Compressed\vspace{2mm}
]{

\includegraphics[width=0.48\textwidth]{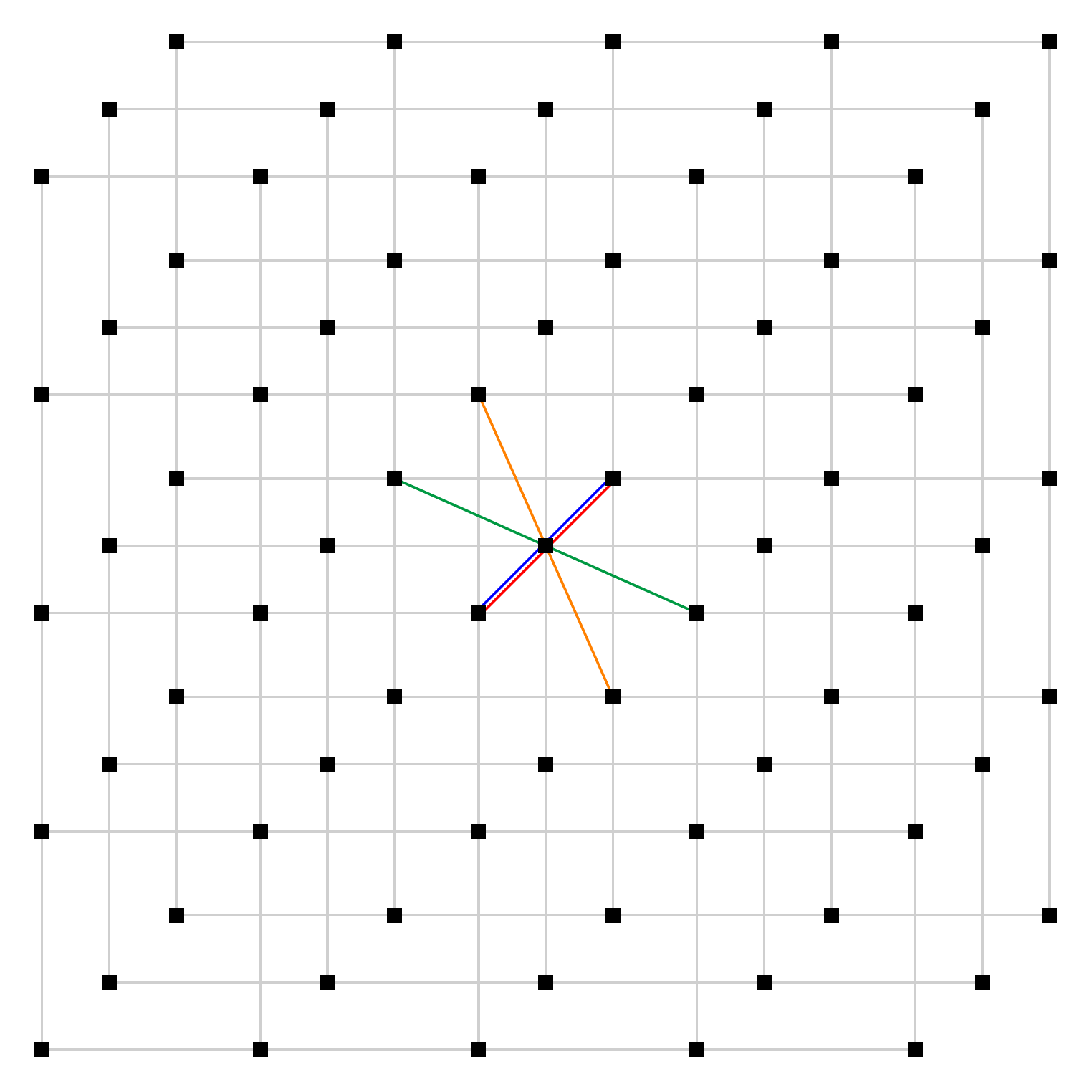}}
\end{figure}
\par\end{center}

\begin{center}
\vfill{}
\par\end{center}

\begin{center}
\begin{figure}[H]
\caption{\centering64 inter-layer edges in color within a $(X,Y,Z)=(5,5,3)$
\emph{tilted} lattice (Pegasus).}
\vspace{15mm}

\subfloat[Tilted classic\vspace{2mm}
]{

\includegraphics[width=0.48\textwidth]{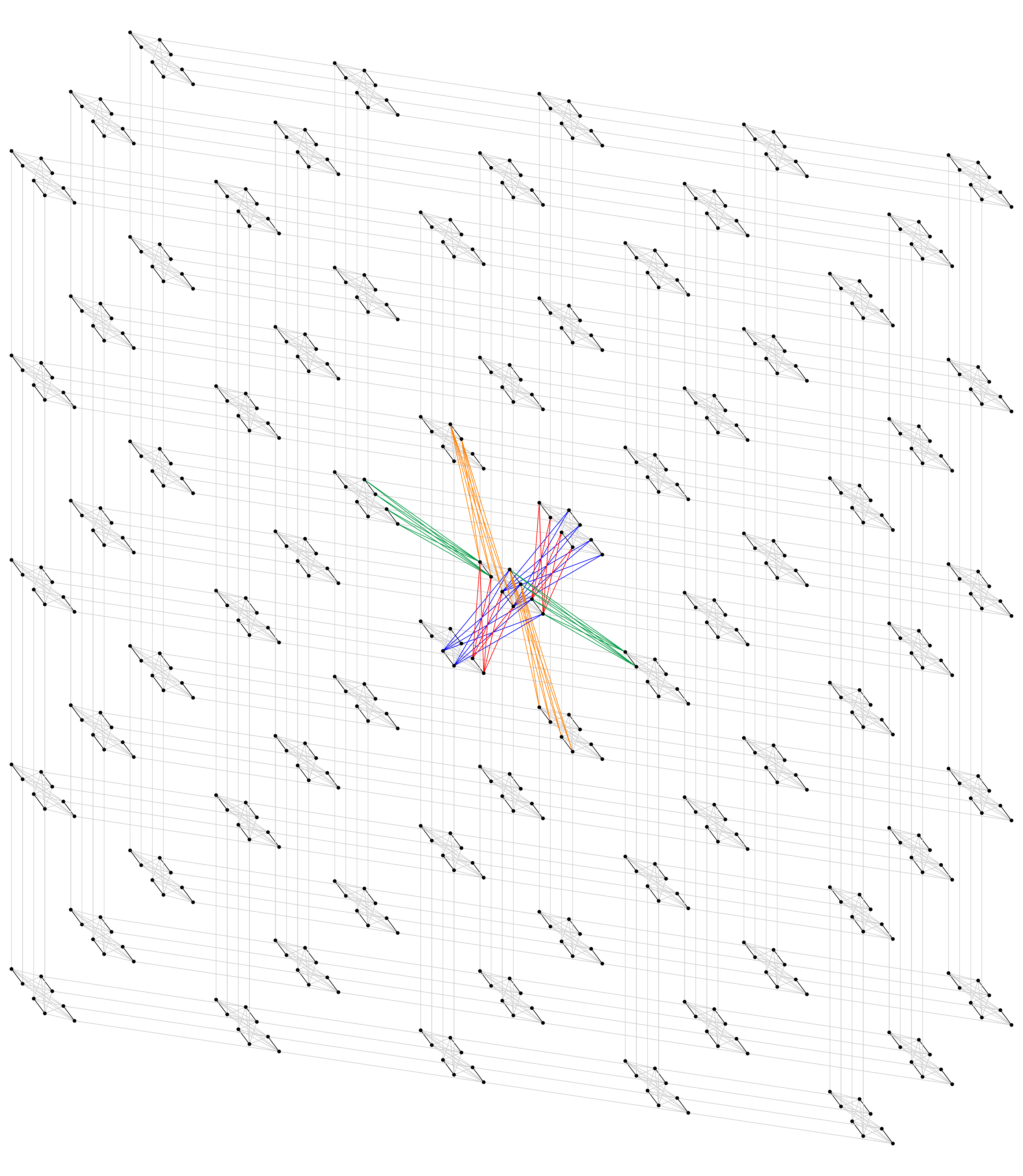}}~~~~~~~~~~\subfloat[Diamond\vspace{2mm}
]{

\includegraphics[width=0.48\textwidth]{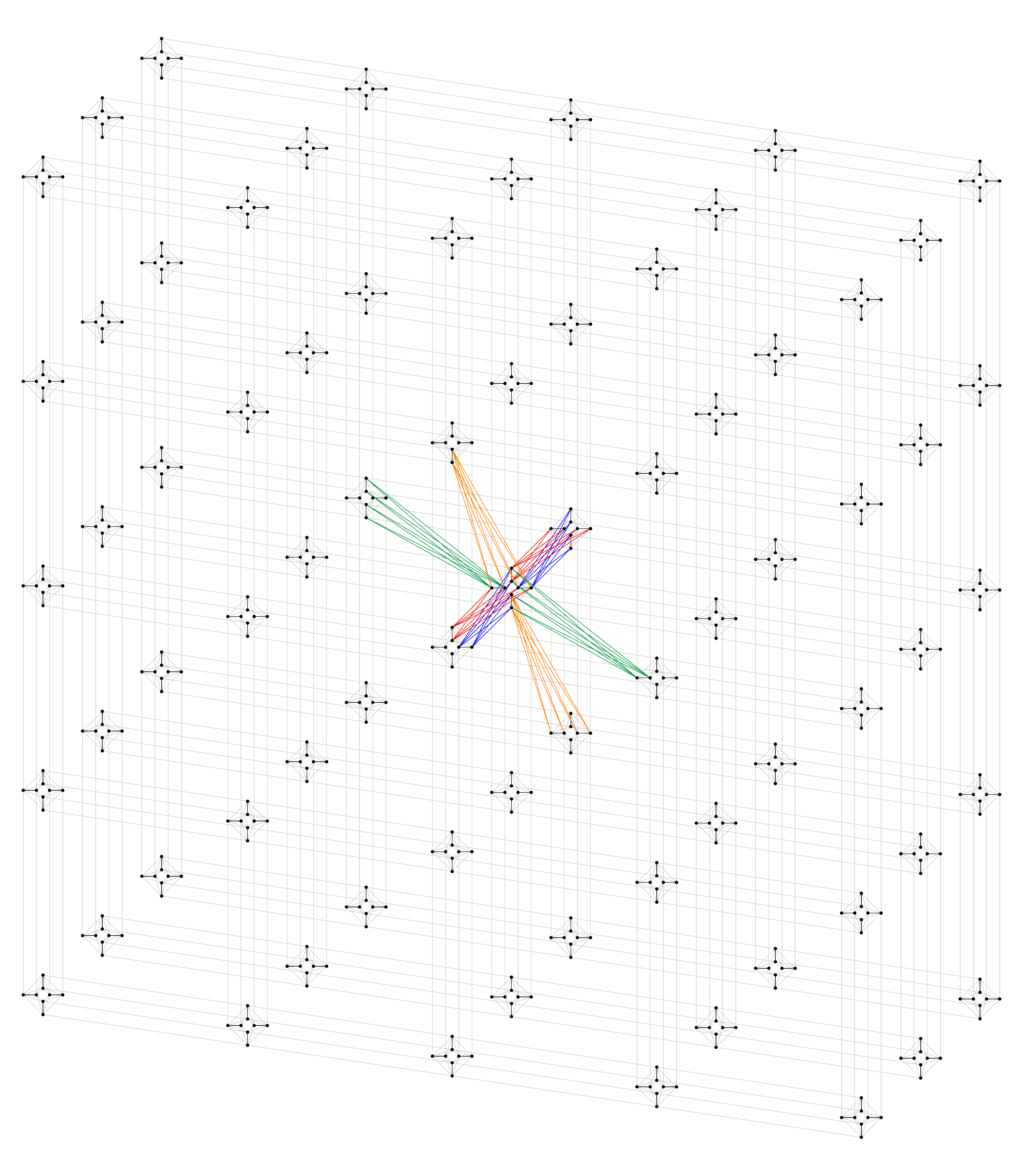}}\vspace{10mm}

\subfloat[Triangle\vspace{2mm}
]{

\includegraphics[width=0.48\textwidth]{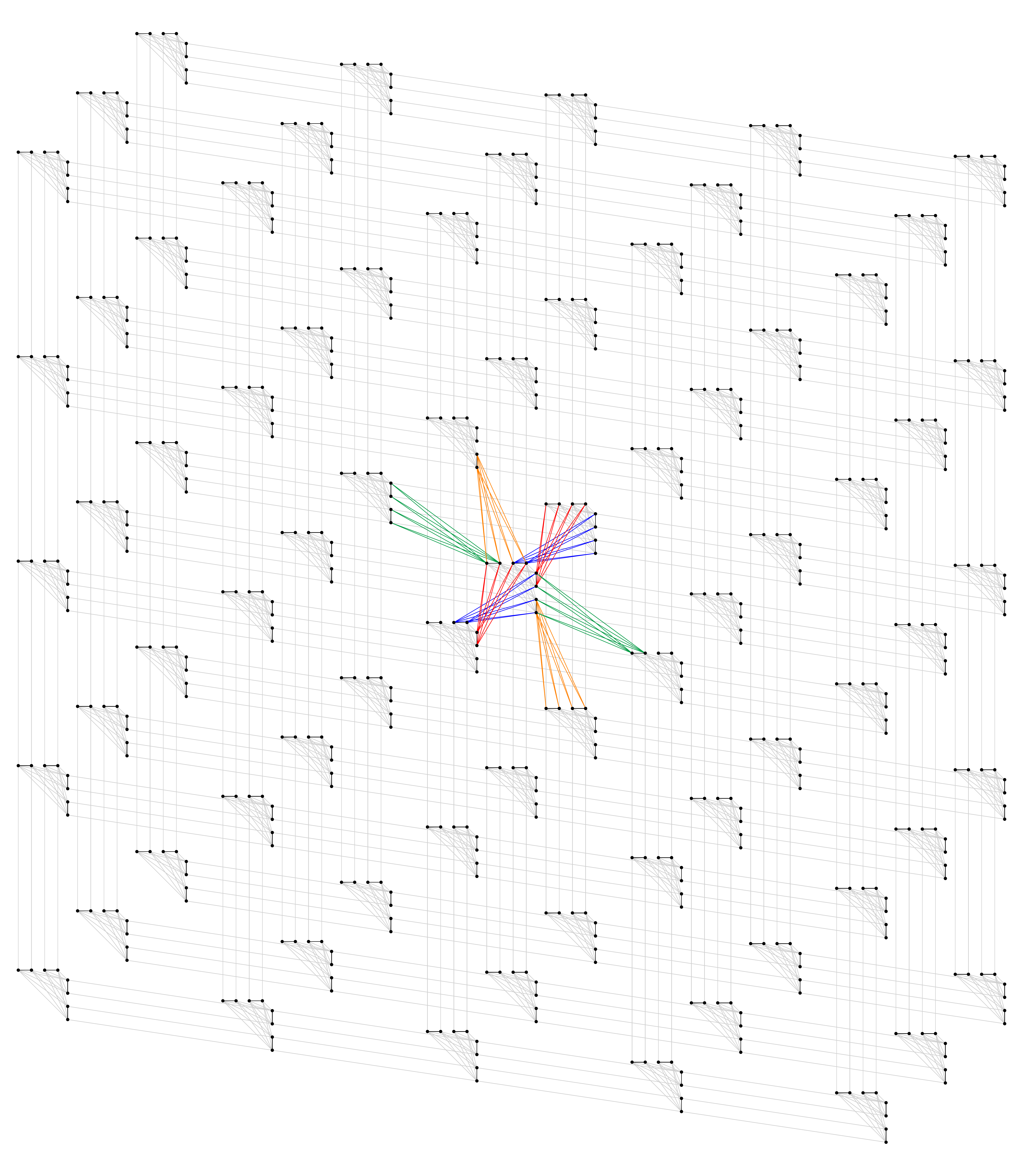}}~~~~~\subfloat[Compressed\vspace{2mm}
]{

\includegraphics[width=0.48\textwidth]{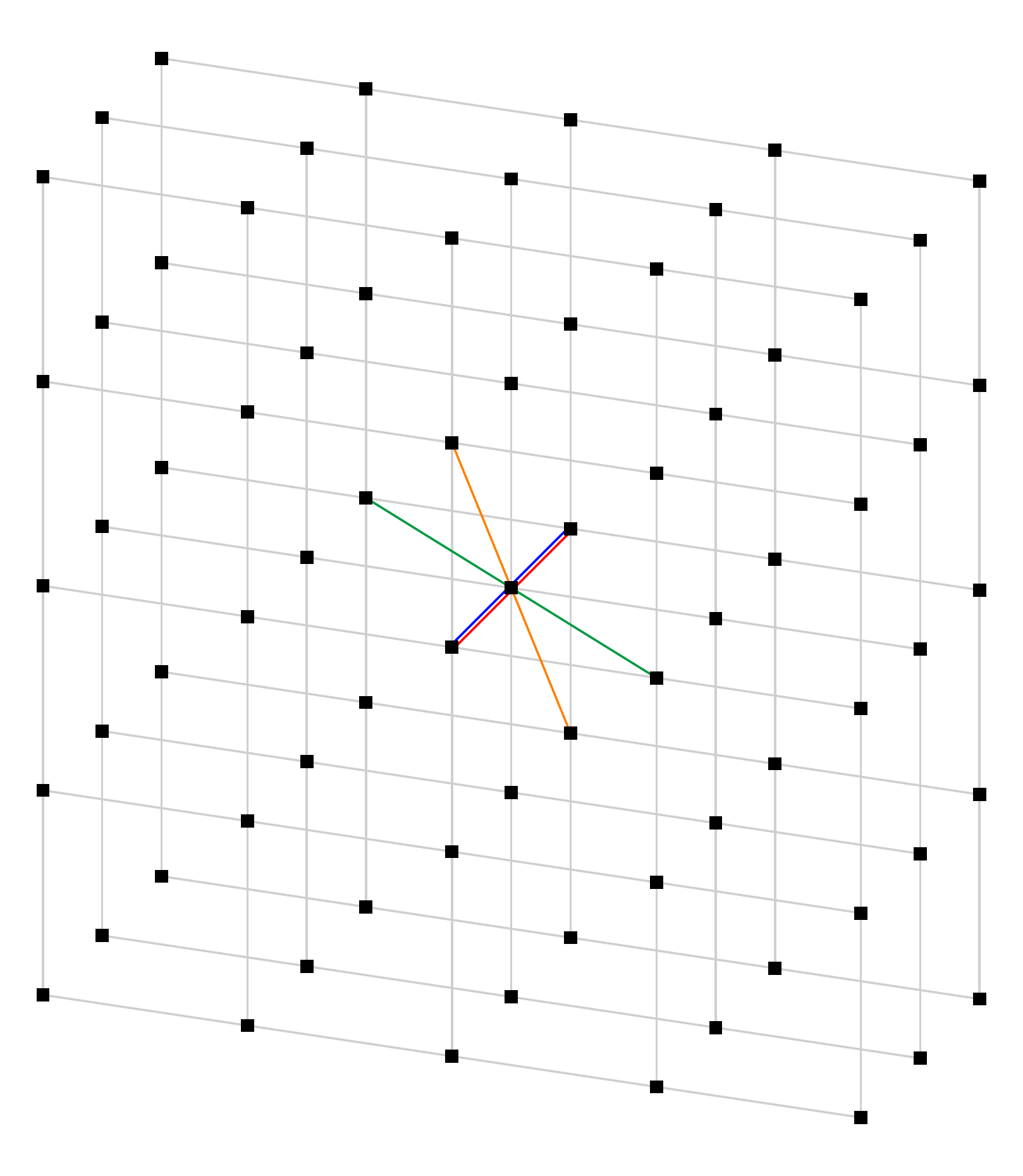}}
\end{figure}
\vfill{}
\begin{figure}[H]
\vspace{10mm}
\caption{\centering$(X,Y,Z)=(2,2,3)$ patch cropped out of Pegasus.}

\vspace{15mm}

\subfloat[Tilted classic\vspace{2mm}
]{

\includegraphics[width=0.48\textwidth]{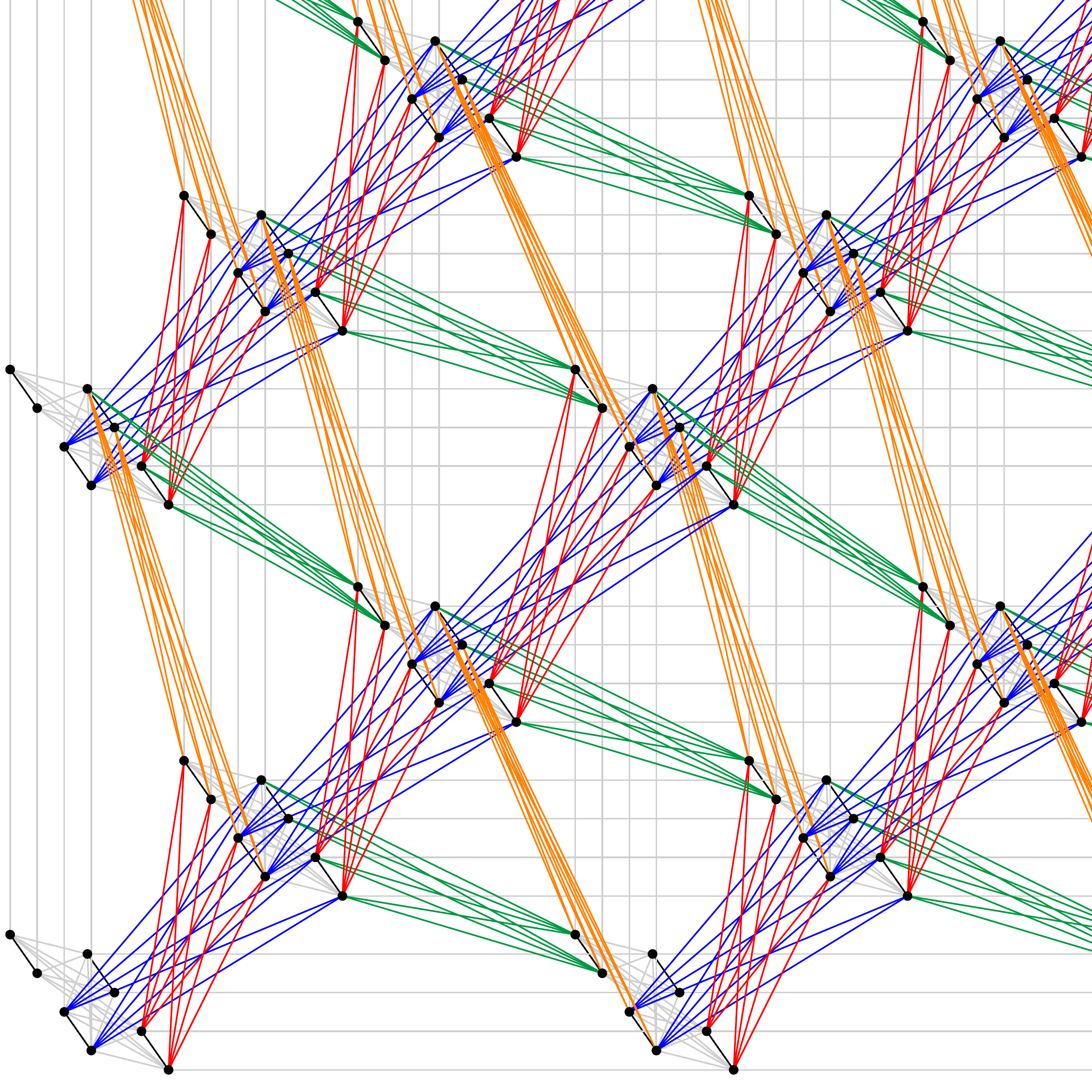}}~~~~~\subfloat[Diamond\vspace{2mm}
]{

\includegraphics[width=0.48\textwidth]{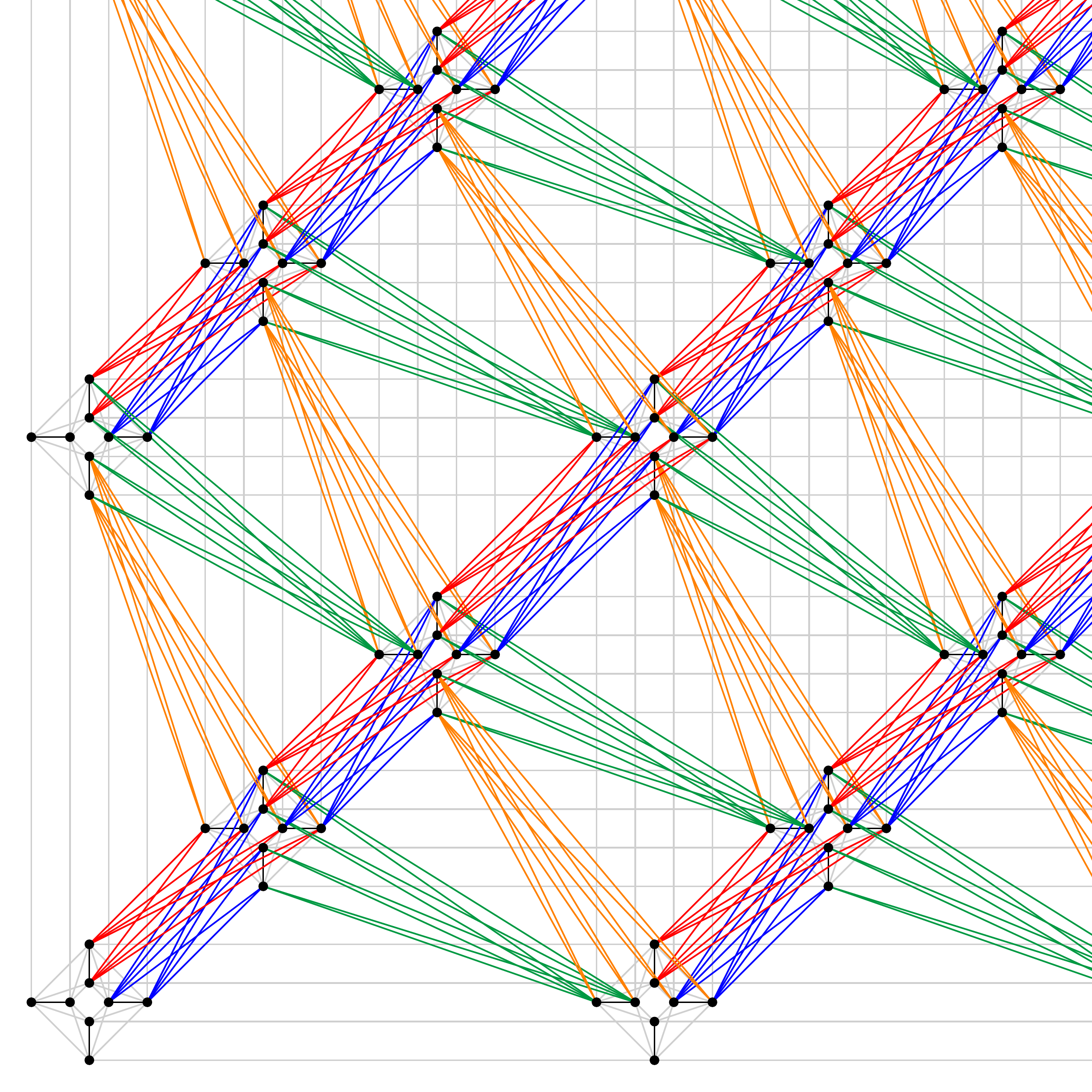}}~~~~~\vspace{15mm}

\subfloat[Triangle\vspace{2mm}
]{

\includegraphics[width=0.48\textwidth]{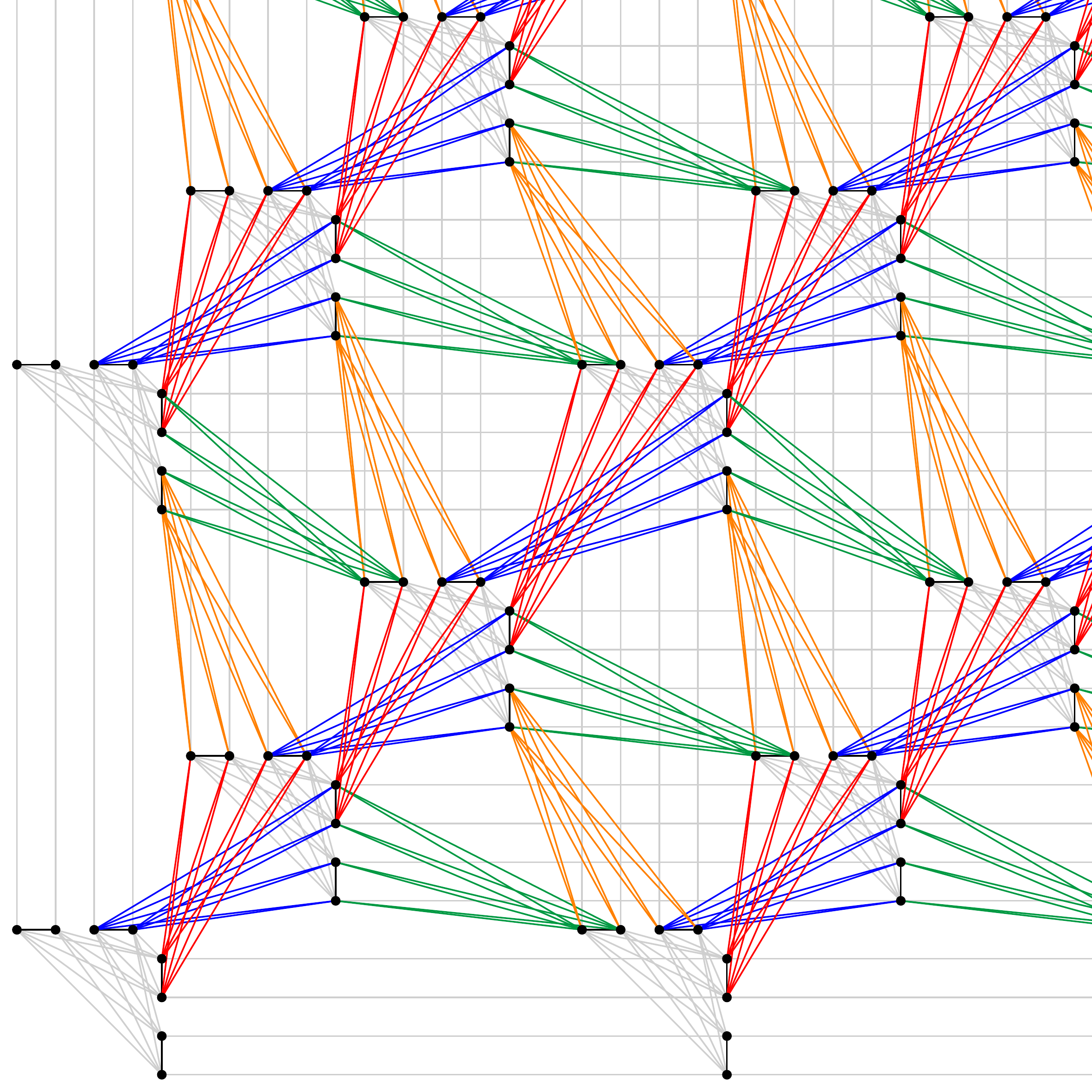}}\subfloat[Compressed\vspace{2mm}
]{

\includegraphics[width=0.48\textwidth]{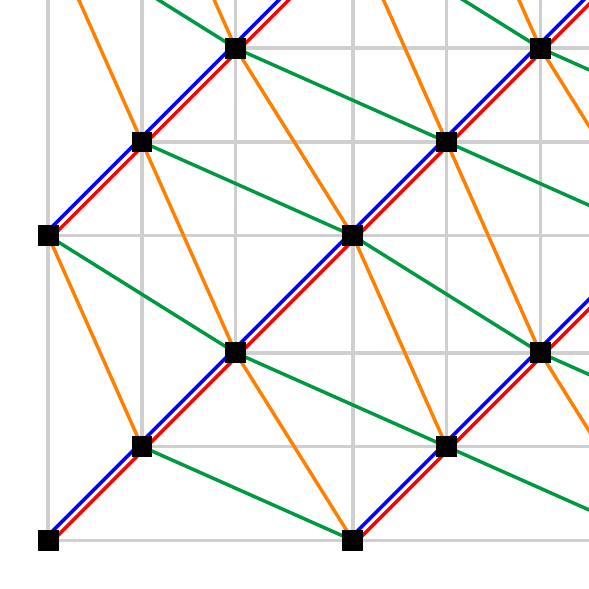}}
\end{figure}
\par\end{center}

\begin{center}
\vfill{}
\par\end{center}

\begin{center}
\begin{figure}[H]
\vspace{10mm}
\caption{\centering$(X,Y,Z)=(2,2,3)$ patch cropped out of a \emph{tilted
}Pegasus.}
\vspace{10mm}

\subfloat[Tilted classic\vspace{2mm}
]{

\includegraphics[width=0.48\textwidth]{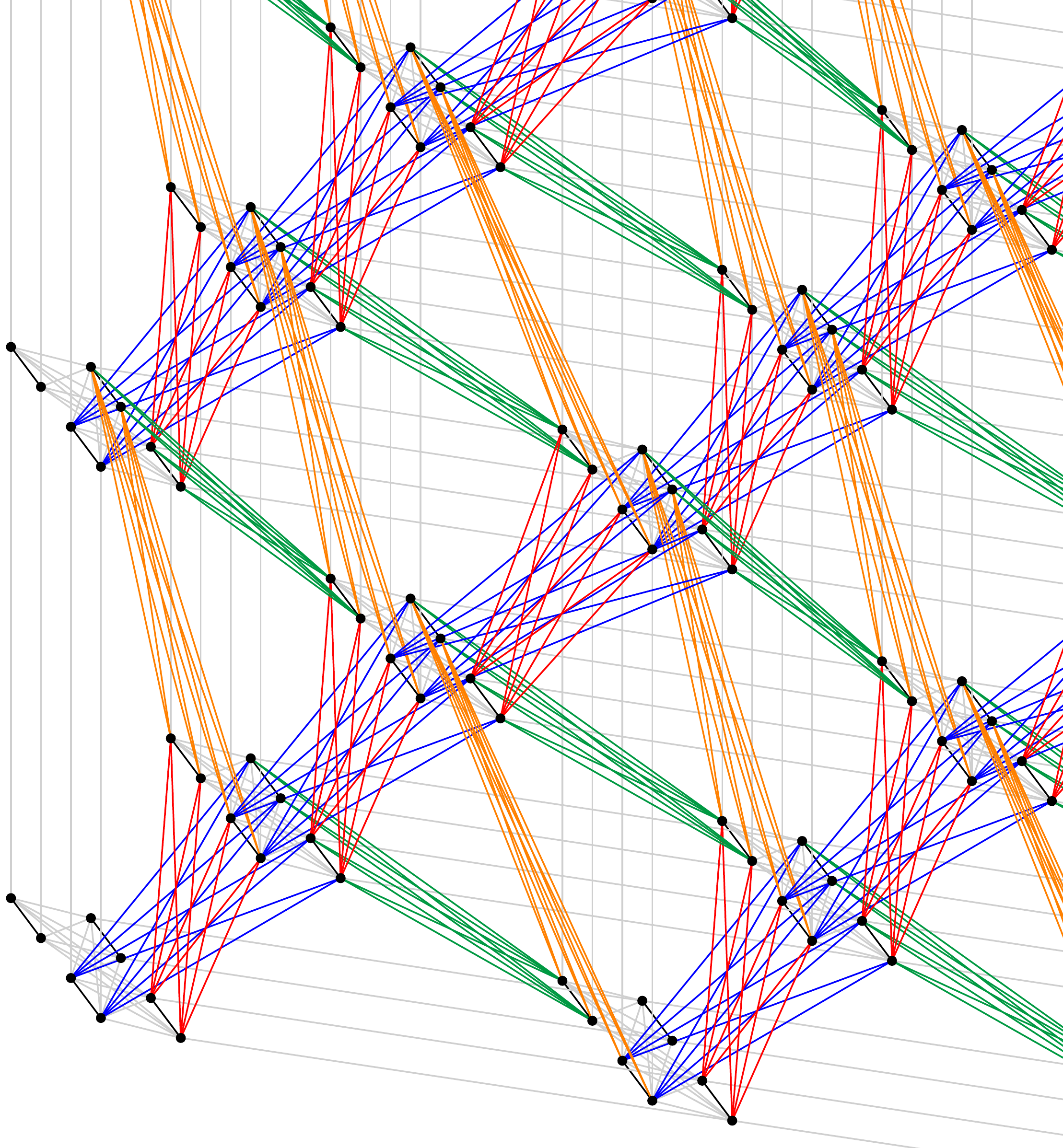}}~~~~~\subfloat[Diamond\vspace{2mm}
]{

\includegraphics[width=0.48\textwidth]{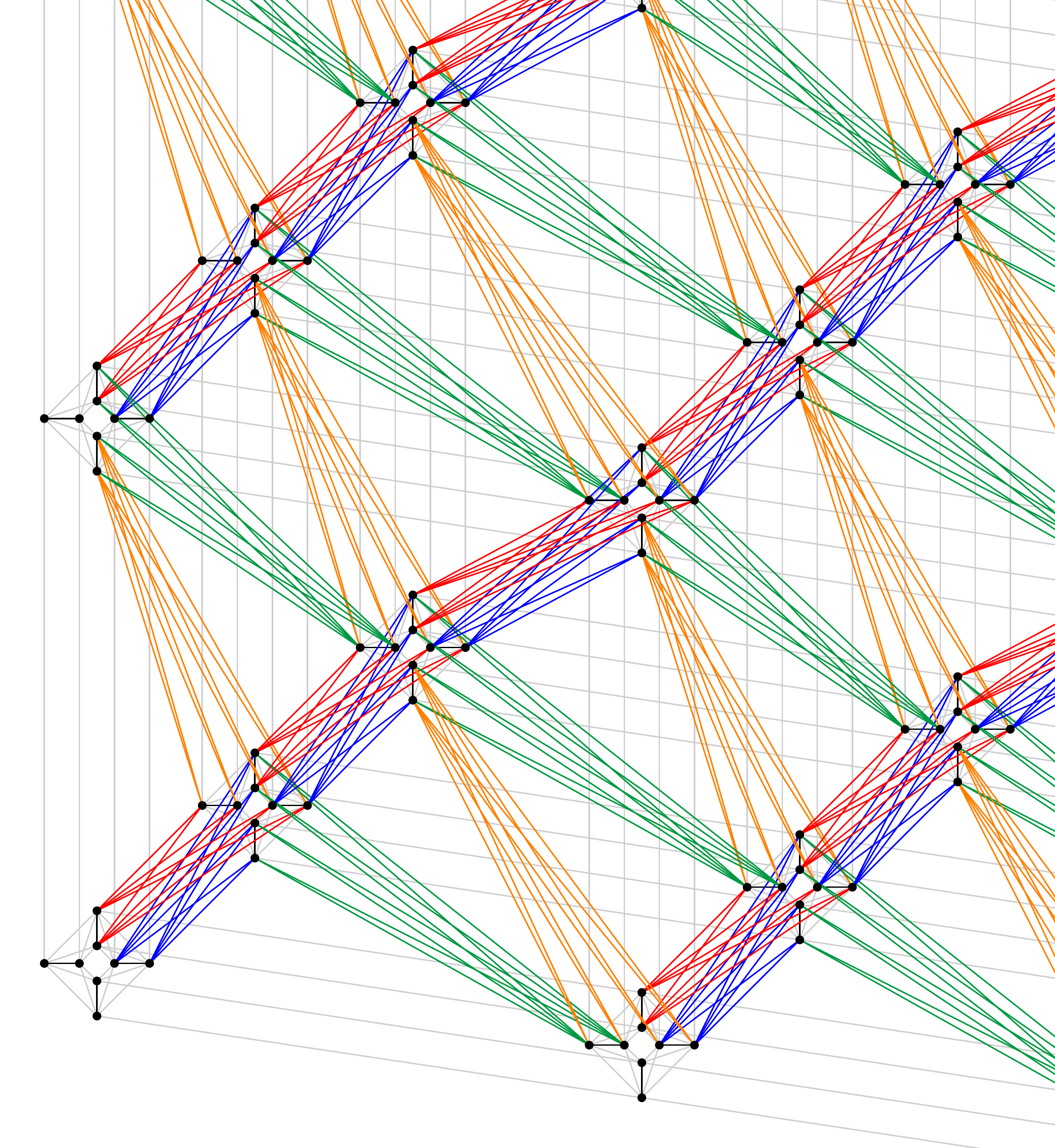}}~~~~~\vspace{10mm}

\subfloat[Triangle\vspace{2mm}
]{

\includegraphics[width=0.48\textwidth]{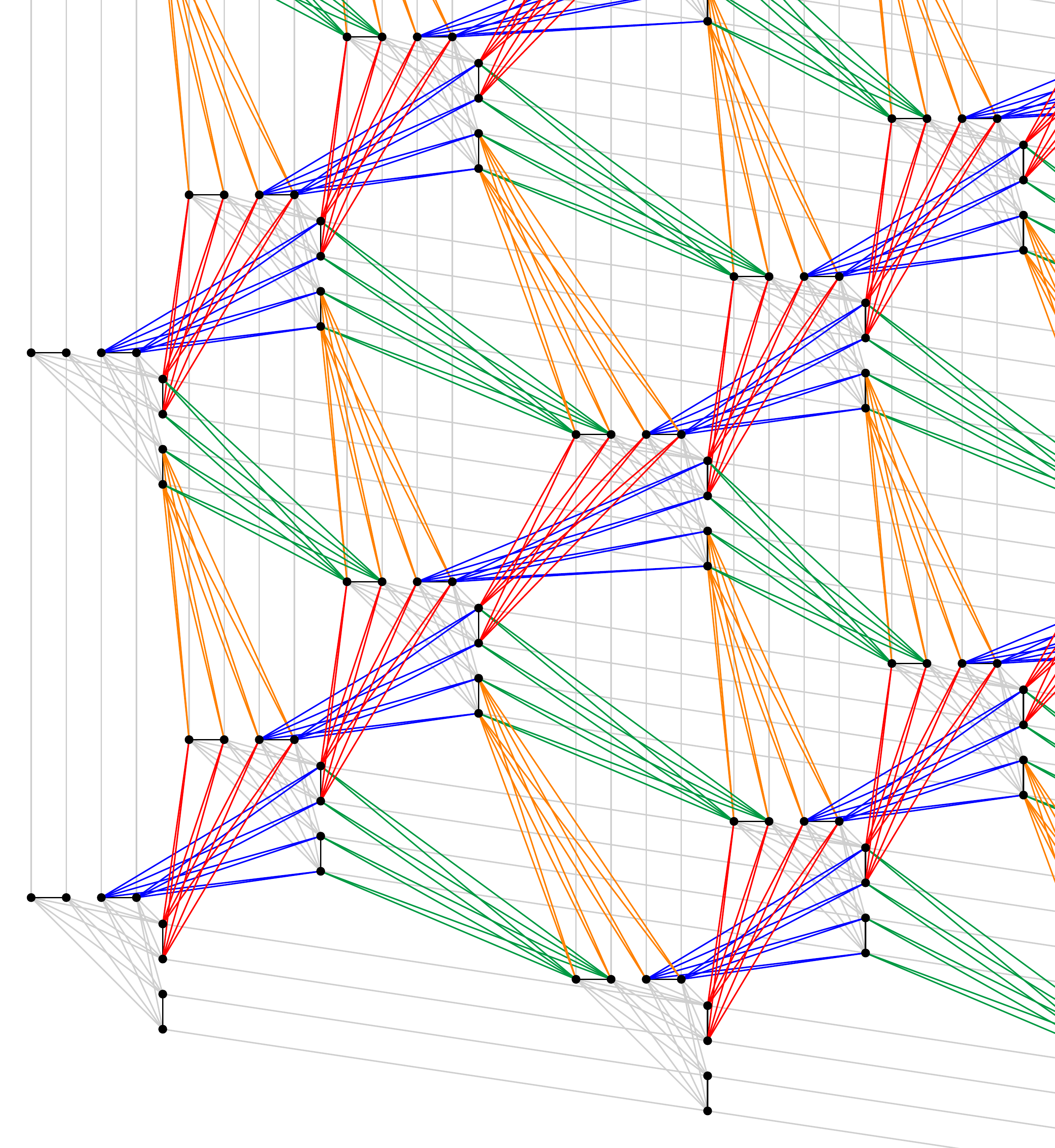}}\subfloat[Compressed\vspace{2mm}
]{

\includegraphics[width=0.48\textwidth]{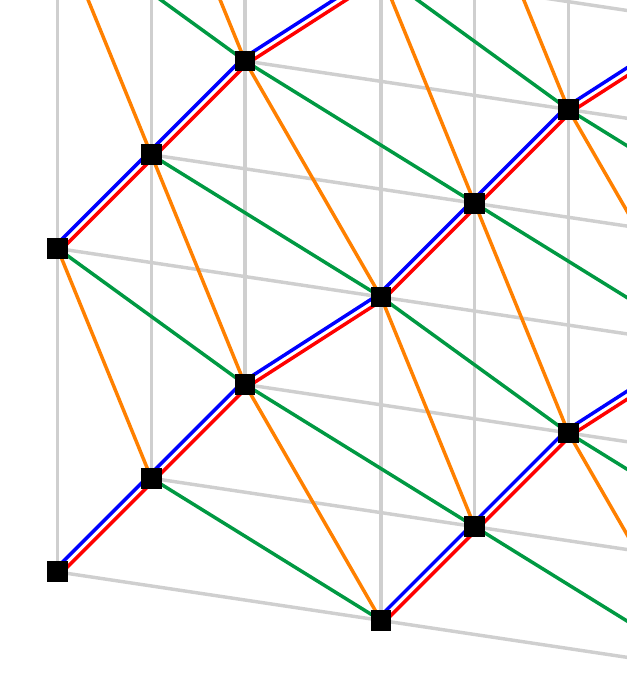}}
\end{figure}
\par\end{center}

\begin{center}
\vfill{}
\begin{figure}[H]
\vspace{10mm}

\caption{\centering$(X,Y,Z)=(2,2,3)$ patch cropped out of Pegasus (with all
Pegasus-only edges either black or light blue).}
\vspace{10mm}

\subfloat[Tilted classic\vspace{2mm}
]{

\includegraphics[width=0.48\textwidth]{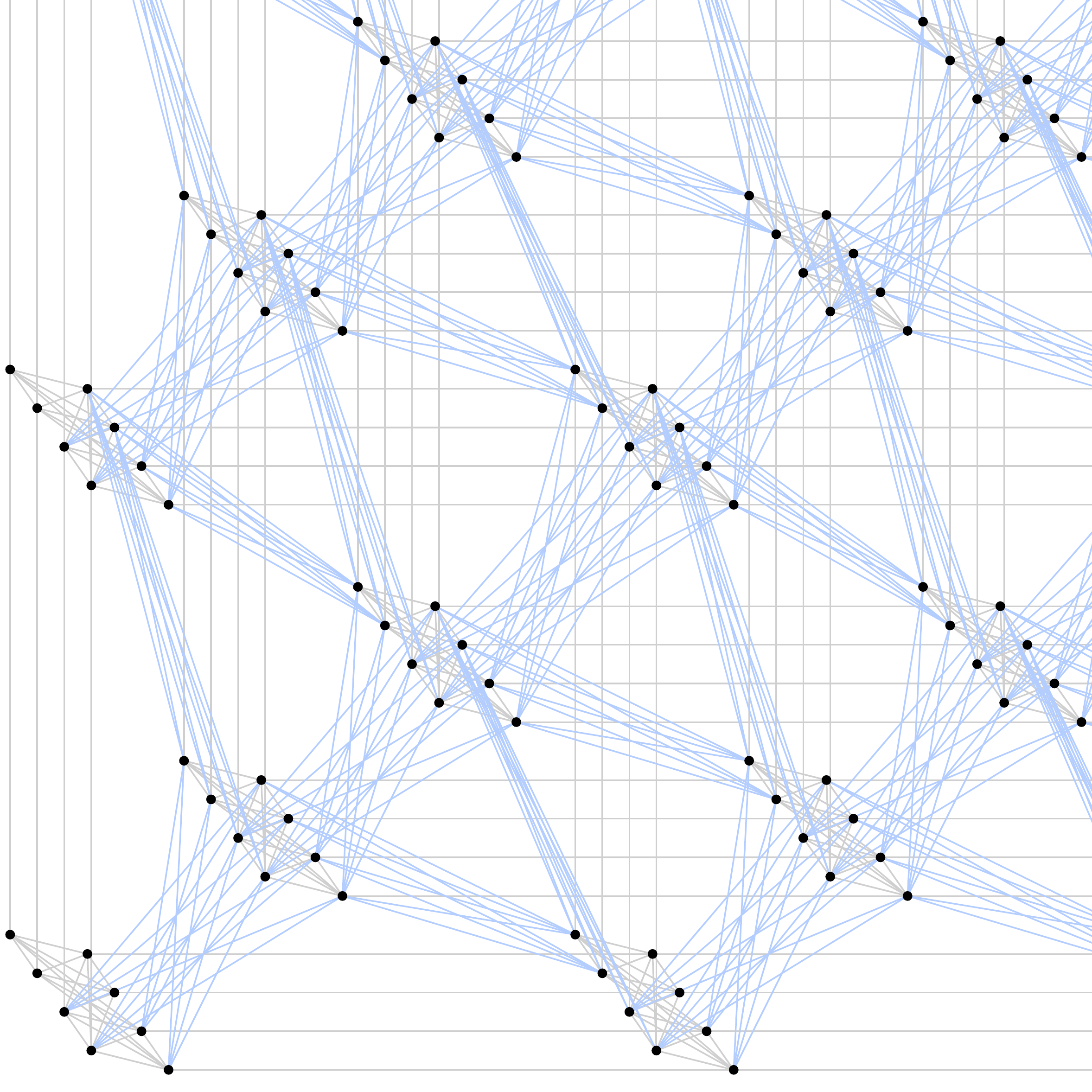}}~~~~~\subfloat[Diamond\vspace{2mm}
]{

\includegraphics[width=0.48\textwidth]{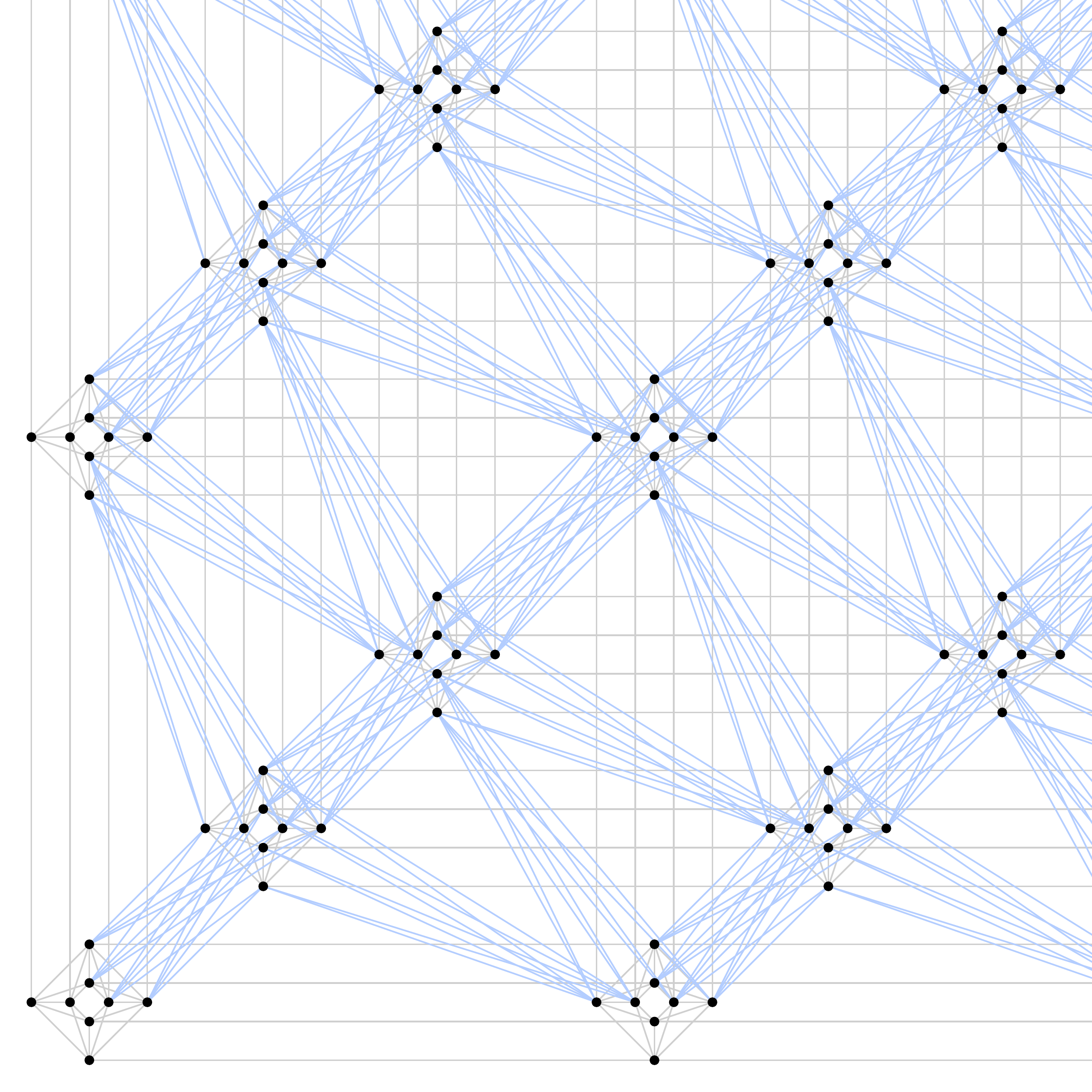}}~~~~~~~~~~\vspace{15mm}

\subfloat[Triangle\vspace{2mm}
]{

\includegraphics[width=0.48\textwidth]{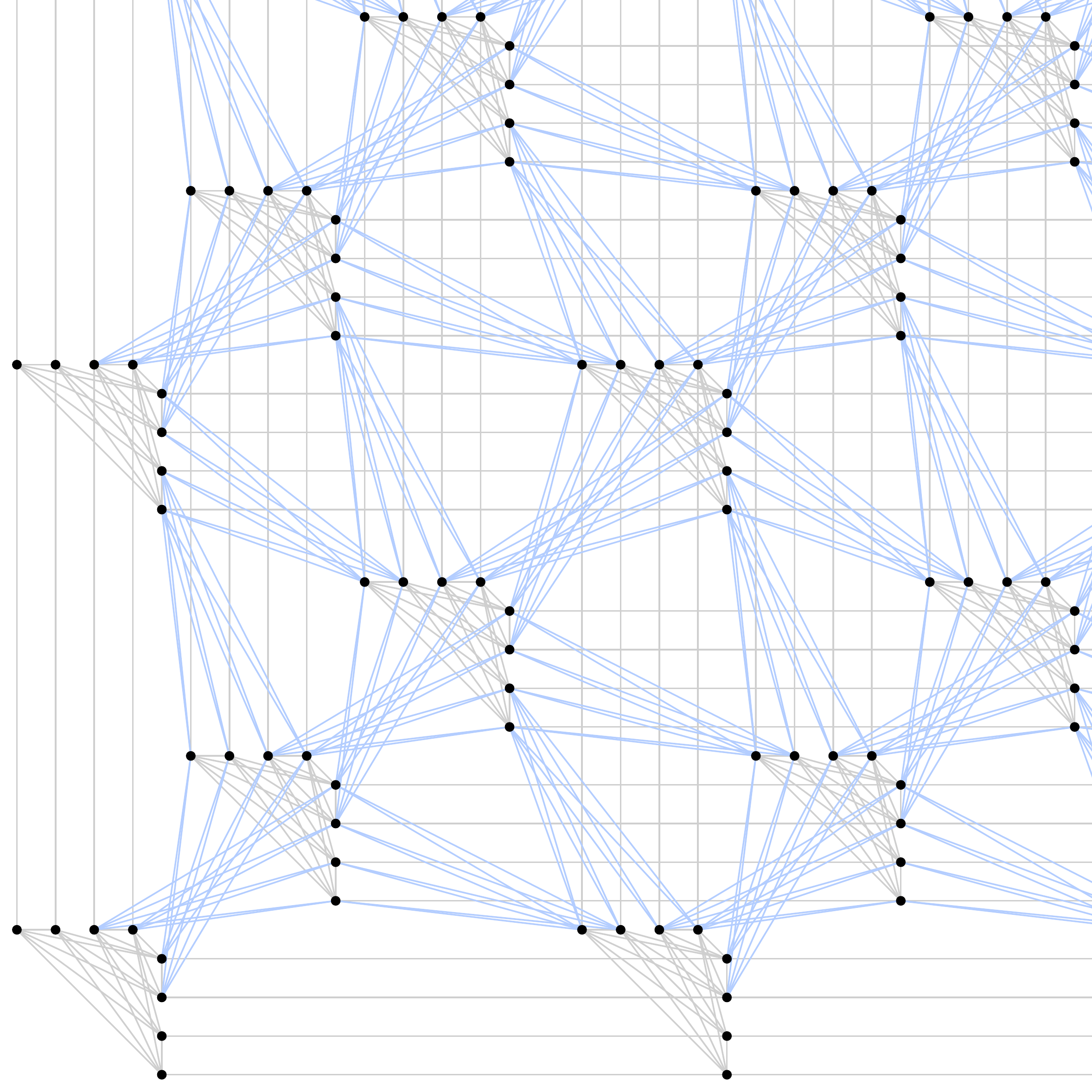}}\subfloat[Compressed\vspace{2mm}
]{

\includegraphics[width=0.48\textwidth]{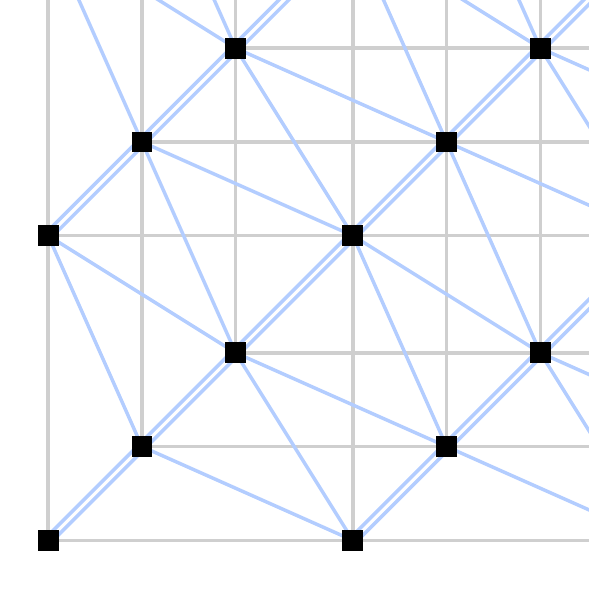}}
\end{figure}
\vfill{}
\par\end{center}

\begin{center}
\begin{figure}[H]
\vspace{10mm}
\caption{\centering$(X,Y,Z)=(2,2,3)$ patch cropped out of \emph{a tilted
}Pegasus (with all Pegasus-only edges either black or light blue).}
\vspace{10mm}

\subfloat[Tilted classic\vspace{2mm}
]{

\includegraphics[width=0.48\textwidth]{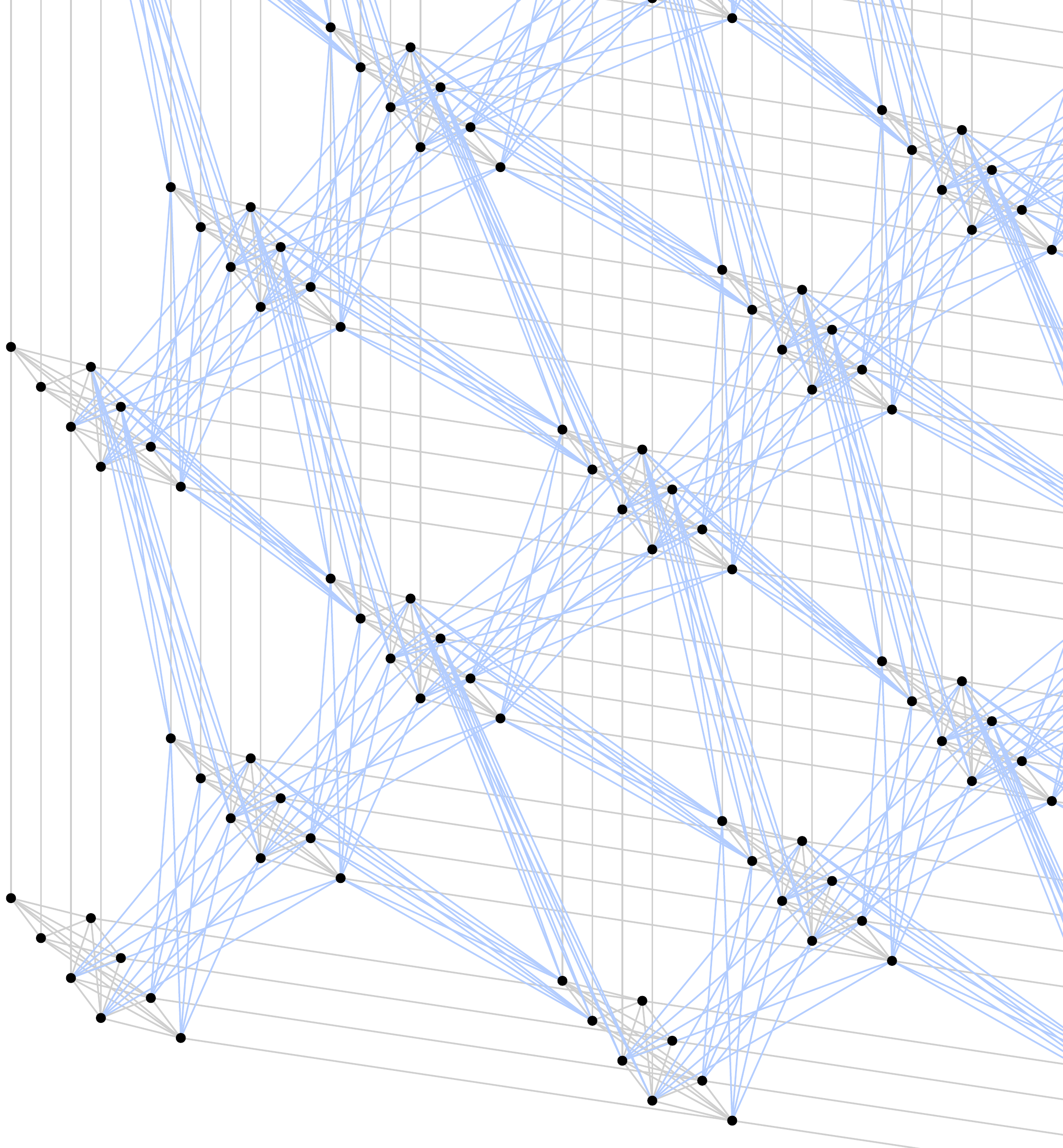}}~~~~~\subfloat[Diamond\vspace{2mm}
]{

\includegraphics[width=0.48\textwidth]{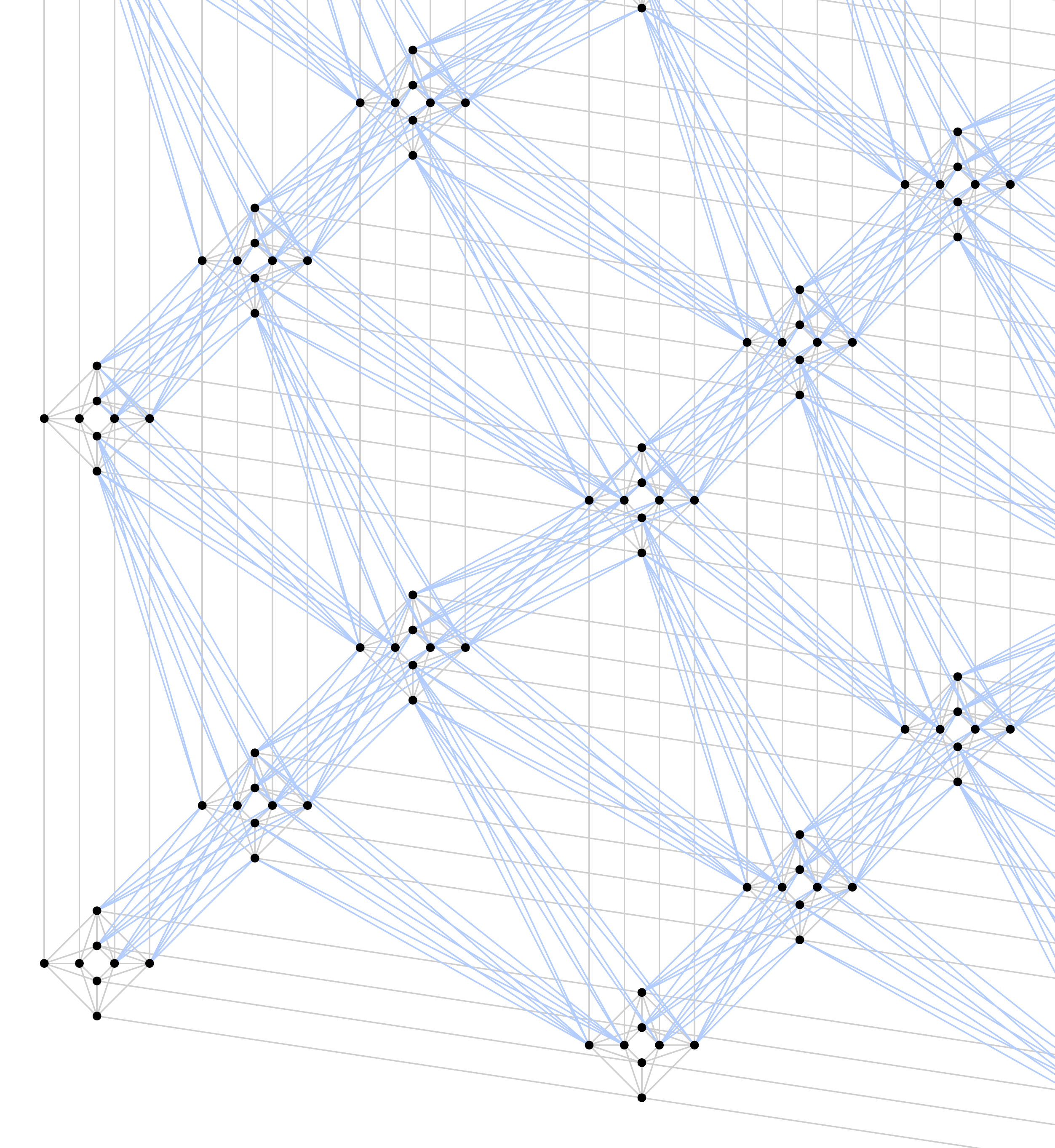}}~~~~~~~~~~\vspace{10mm}

\subfloat[Triangle\vspace{2mm}
]{

\includegraphics[width=0.48\textwidth]{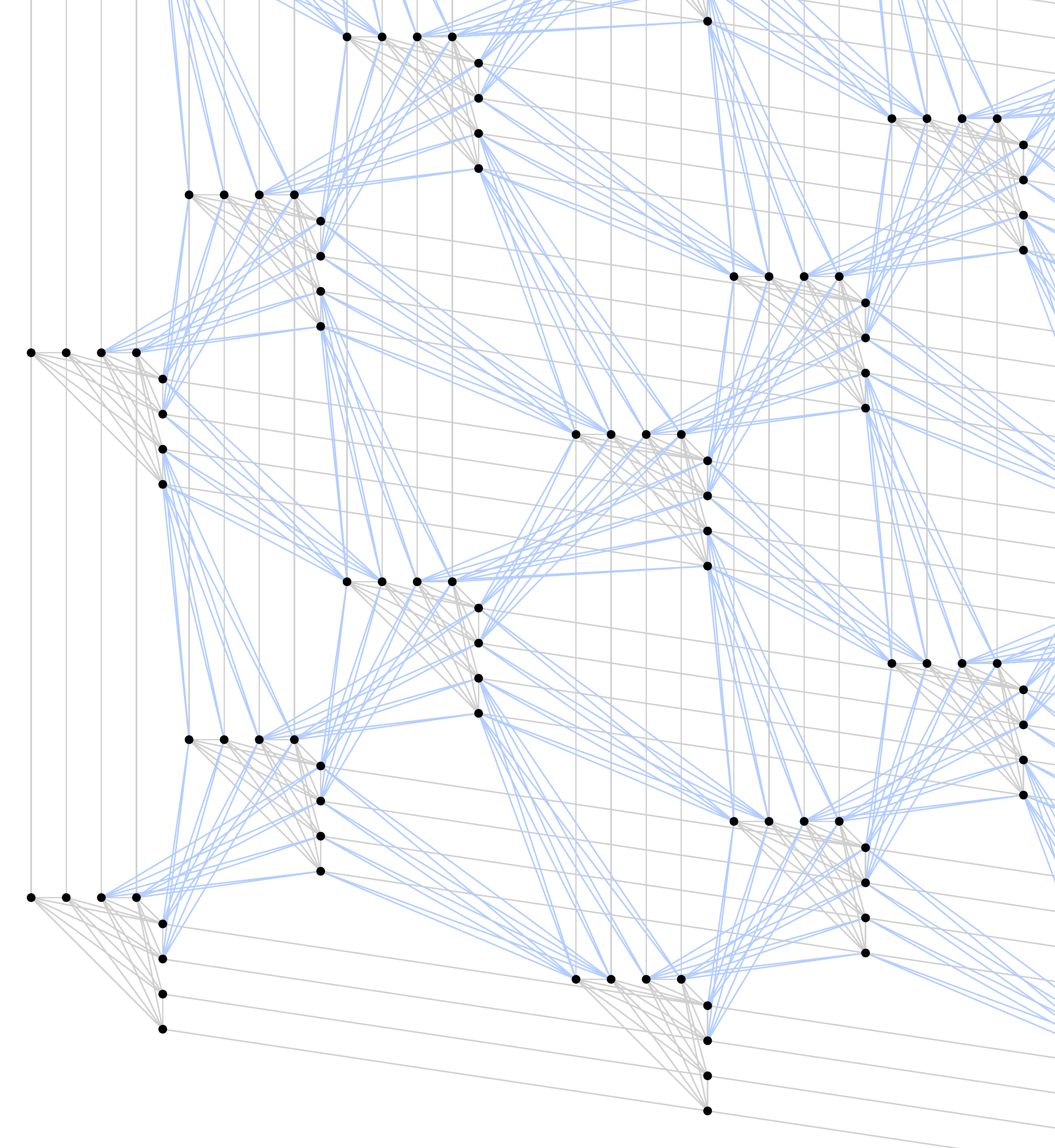}}\subfloat[Compressed\vspace{2mm}
]{

\includegraphics[width=0.48\textwidth]{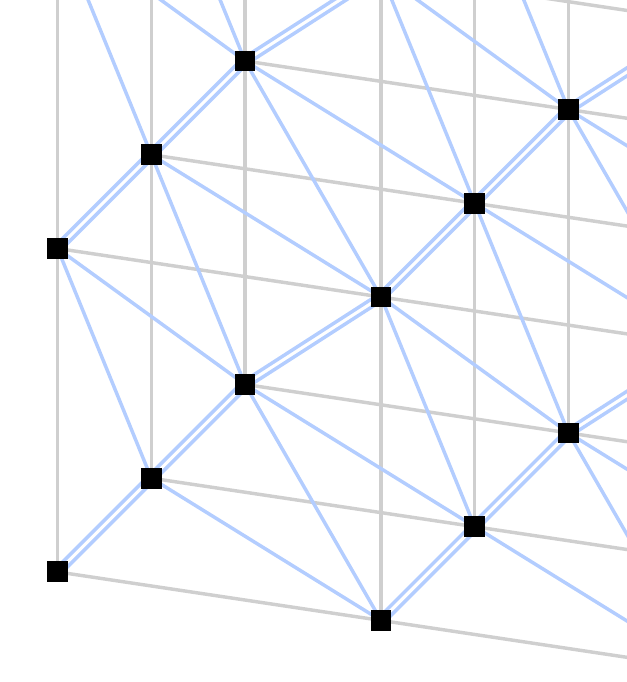}}
\end{figure}
\vfill{}
\begin{figure}[H]
\vspace{10mm}
\caption{\centering$(X,Y,Z)=(5,5,3)$ lattice of Pegasus.}
\vspace{10mm}

\subfloat[Tilted classic\vspace{2mm}
]{

\includegraphics[width=0.48\textwidth]{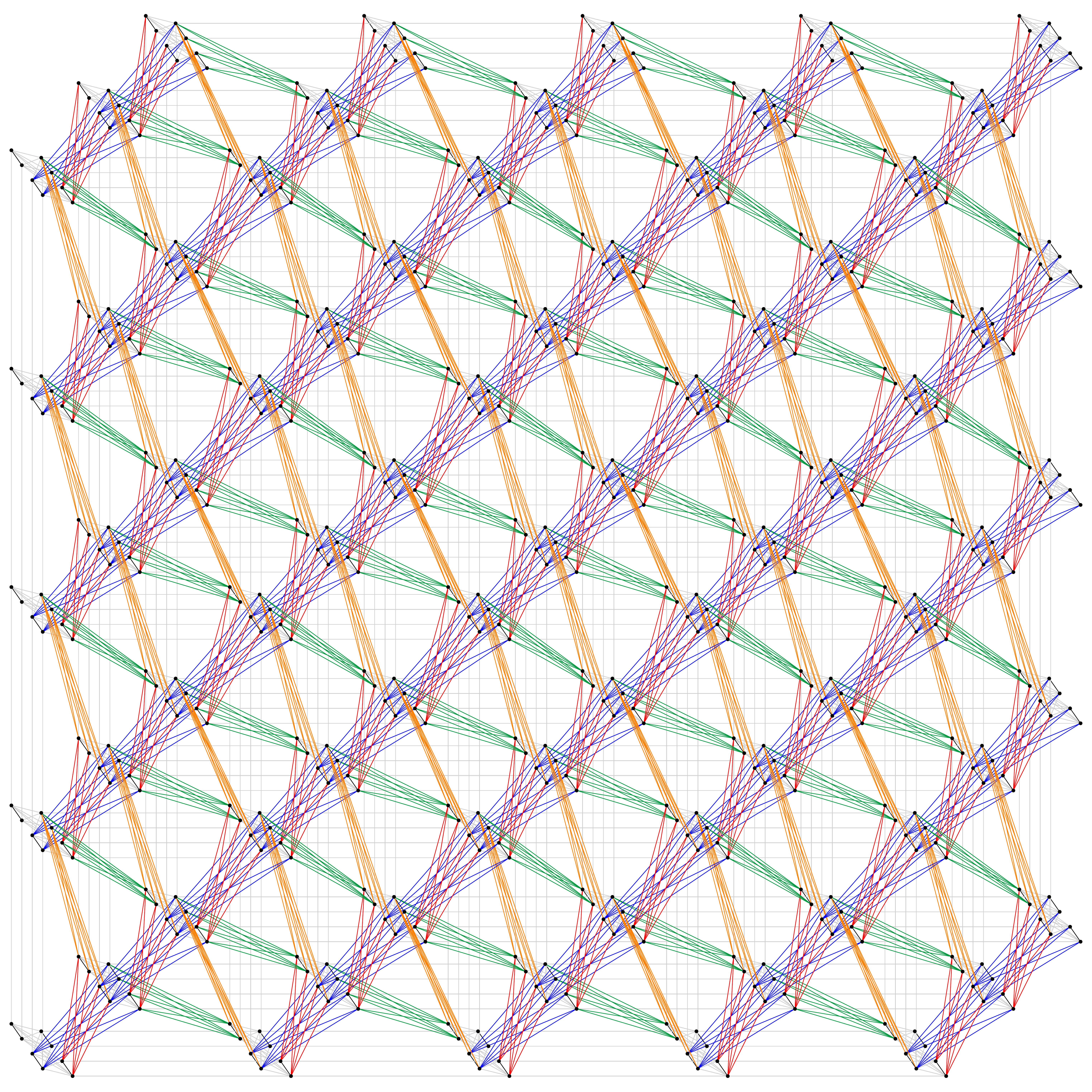}}~~~~~\subfloat[Diamond\vspace{2mm}
]{

\includegraphics[width=0.48\textwidth]{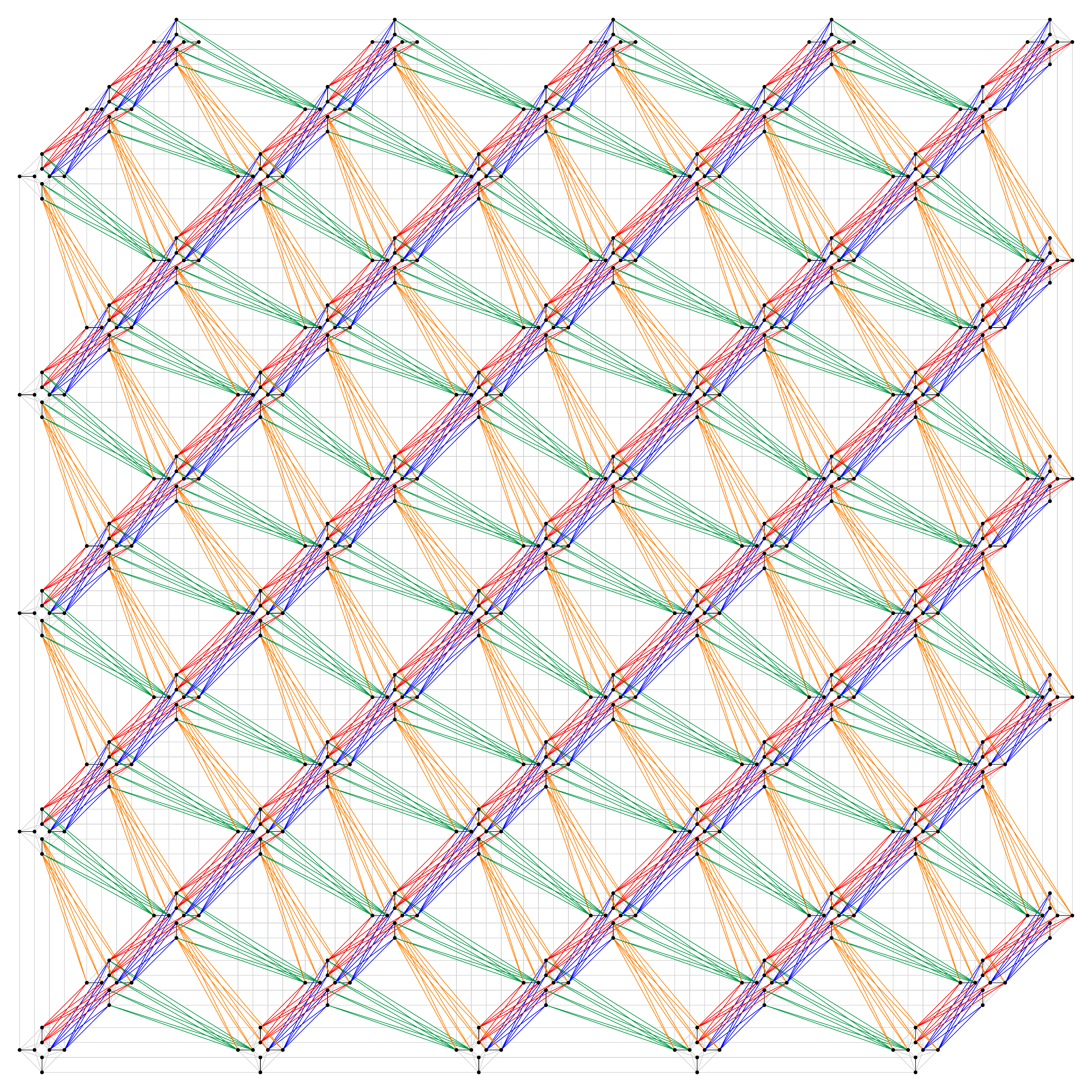}}~~~~~~~~~~\vspace{10mm}

\subfloat[Triangle\vspace{2mm}
]{

\includegraphics[width=0.48\textwidth]{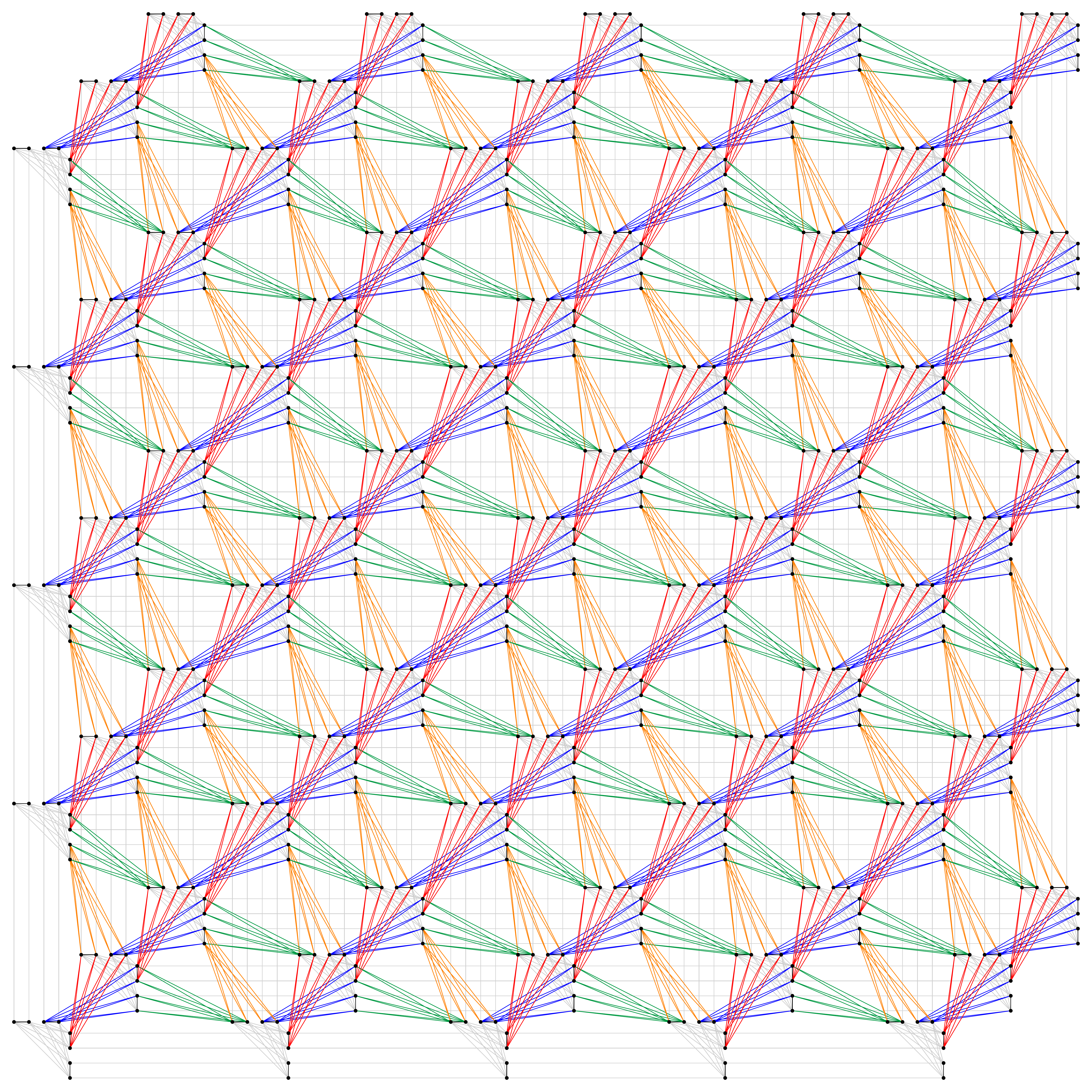}}~\subfloat[Compressed\vspace{2mm}
]{

\includegraphics[width=0.48\textwidth]{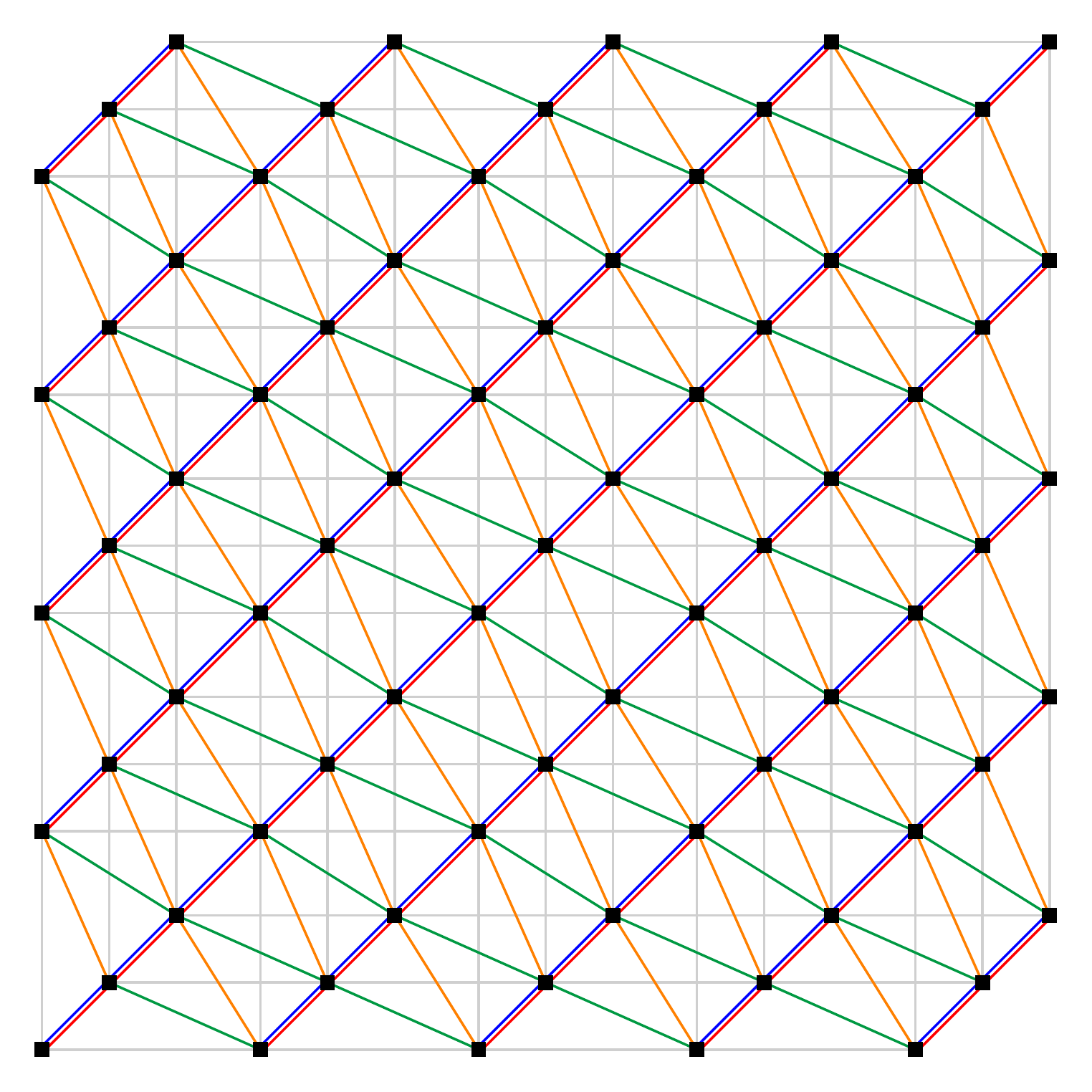}}
\end{figure}
\vfill{}
\par\end{center}

\begin{center}
\begin{figure}[H]
\vspace{10mm}
\caption{\centering$(X,Y,Z)=(5,5,3)$ \emph{tilted} lattice of Pegasus.}

\vspace{10mm}

\subfloat[Tilted classic\vspace{2mm}
]{

\includegraphics[width=0.48\textwidth]{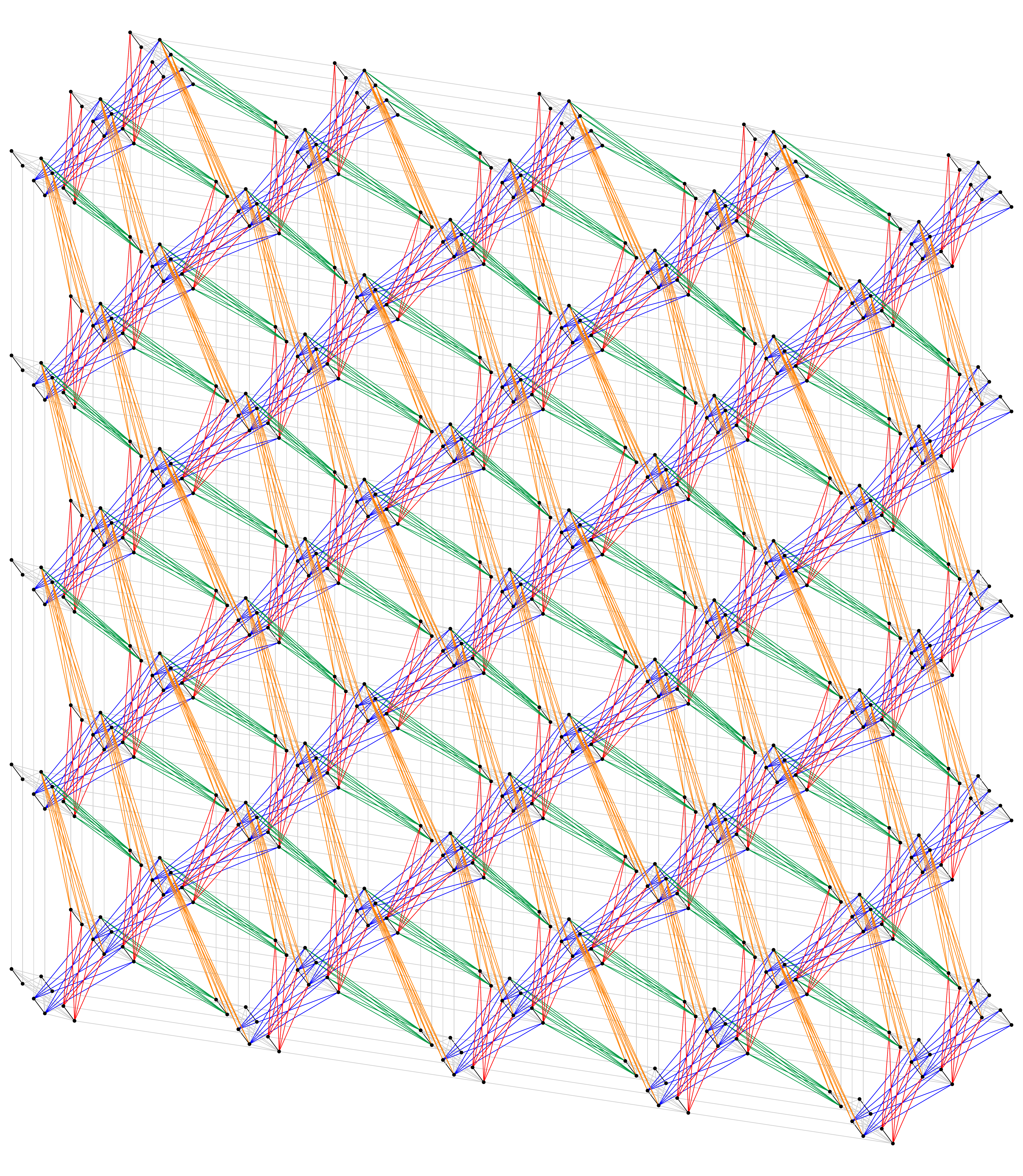}}~~~~~~\subfloat[Diamond\vspace{2mm}
]{

\includegraphics[width=0.48\textwidth]{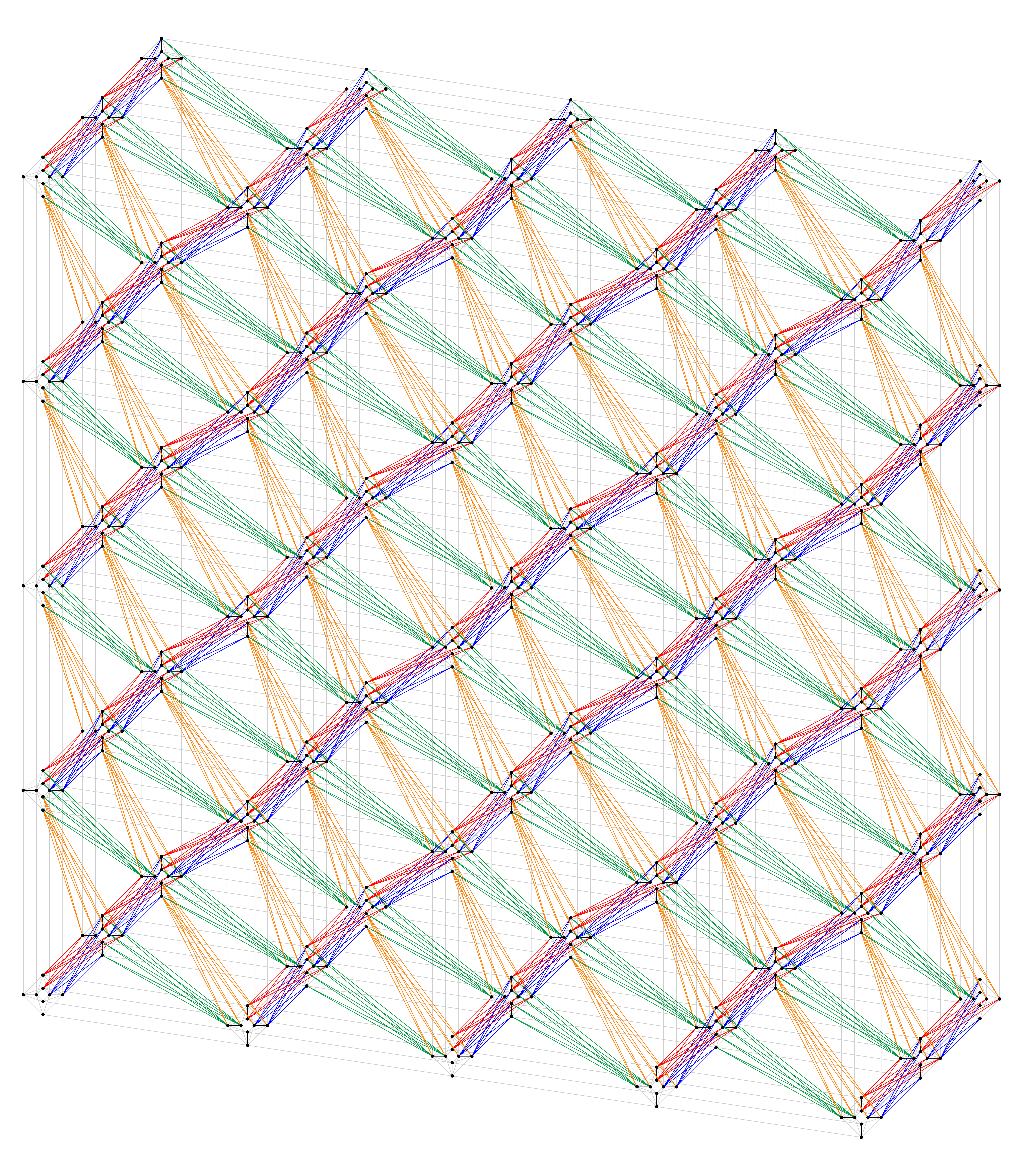}}~~~~~~~~~~\vspace{7mm}

\subfloat[Triangle\vspace{2mm}
]{

\includegraphics[width=0.48\textwidth]{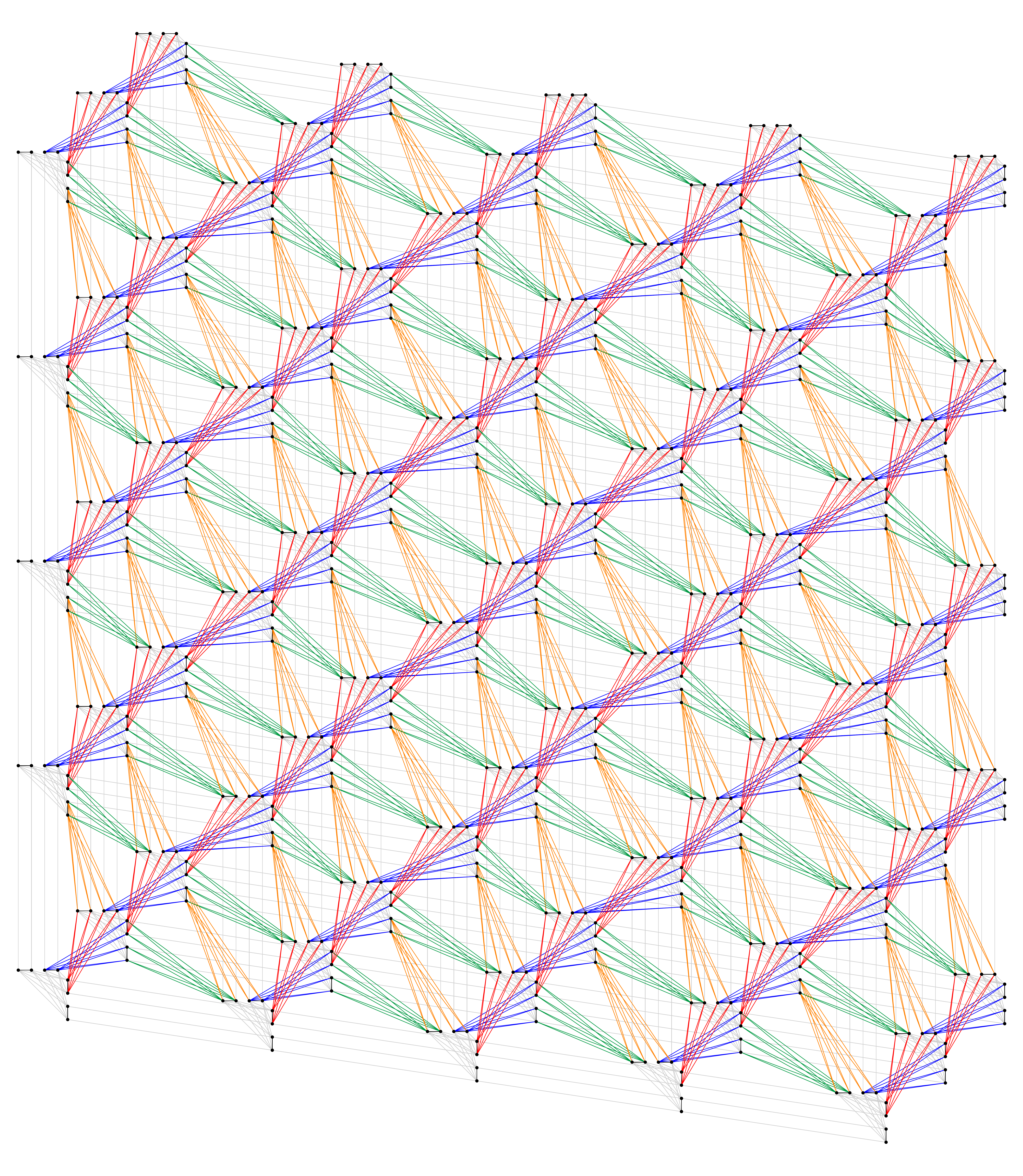}}~~~\subfloat[Compressed\vspace{2mm}
]{

\includegraphics[width=0.48\textwidth]{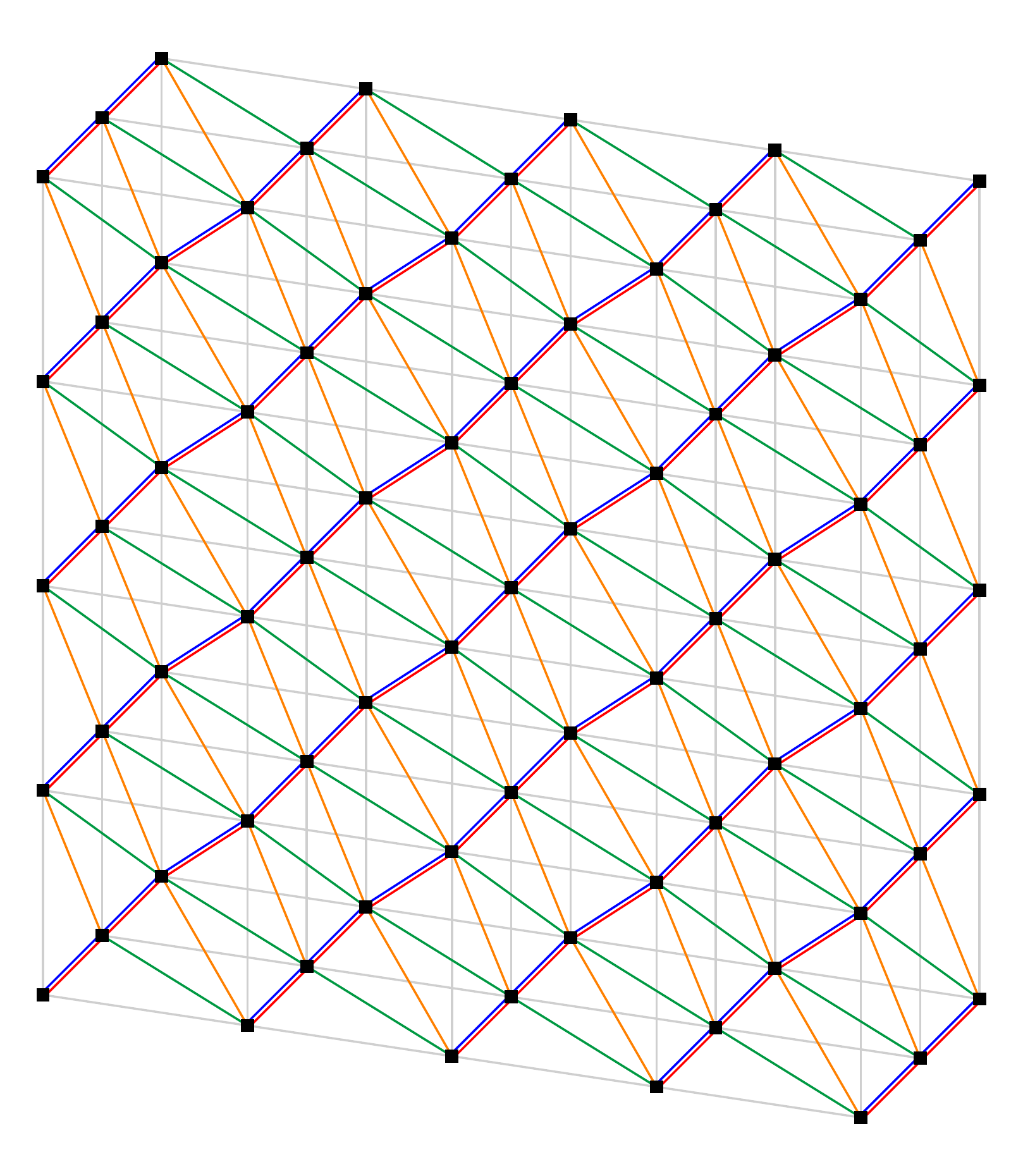}}
\end{figure}
\vfill{}
\par\end{center}

\begin{center}
\begin{figure}[H]
\vspace{10mm}
\caption{\centering$(X,Y,Z)=(5,5,3)$ lattice of Pegasus (with all Pegasus-only
edges either black or light blue).}
\vspace{10mm}

\subfloat[Tilted classic\vspace{2mm}
]{

\includegraphics[width=0.48\textwidth]{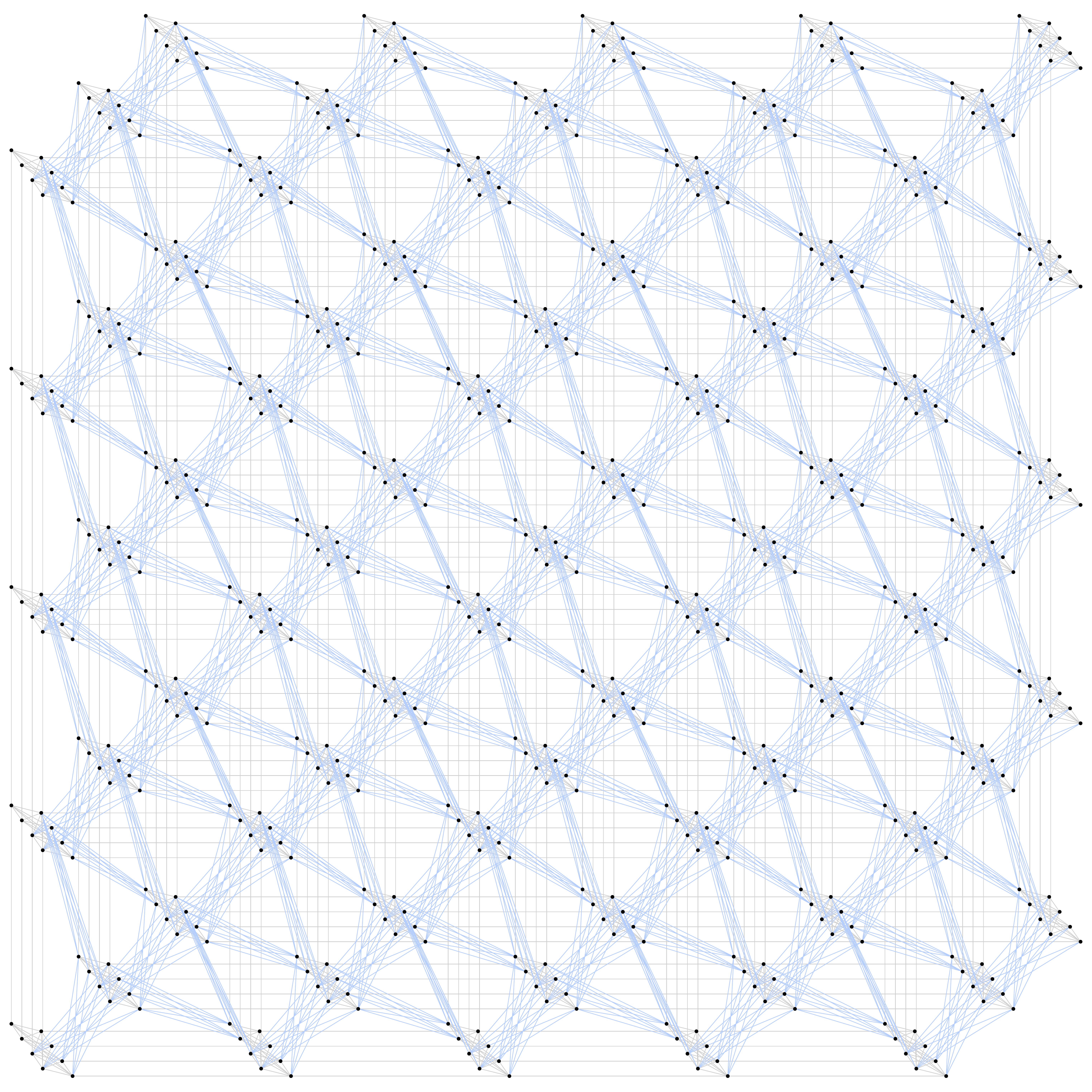}}~~~~~~\subfloat[Diamond\vspace{2mm}
]{

\includegraphics[width=0.48\textwidth]{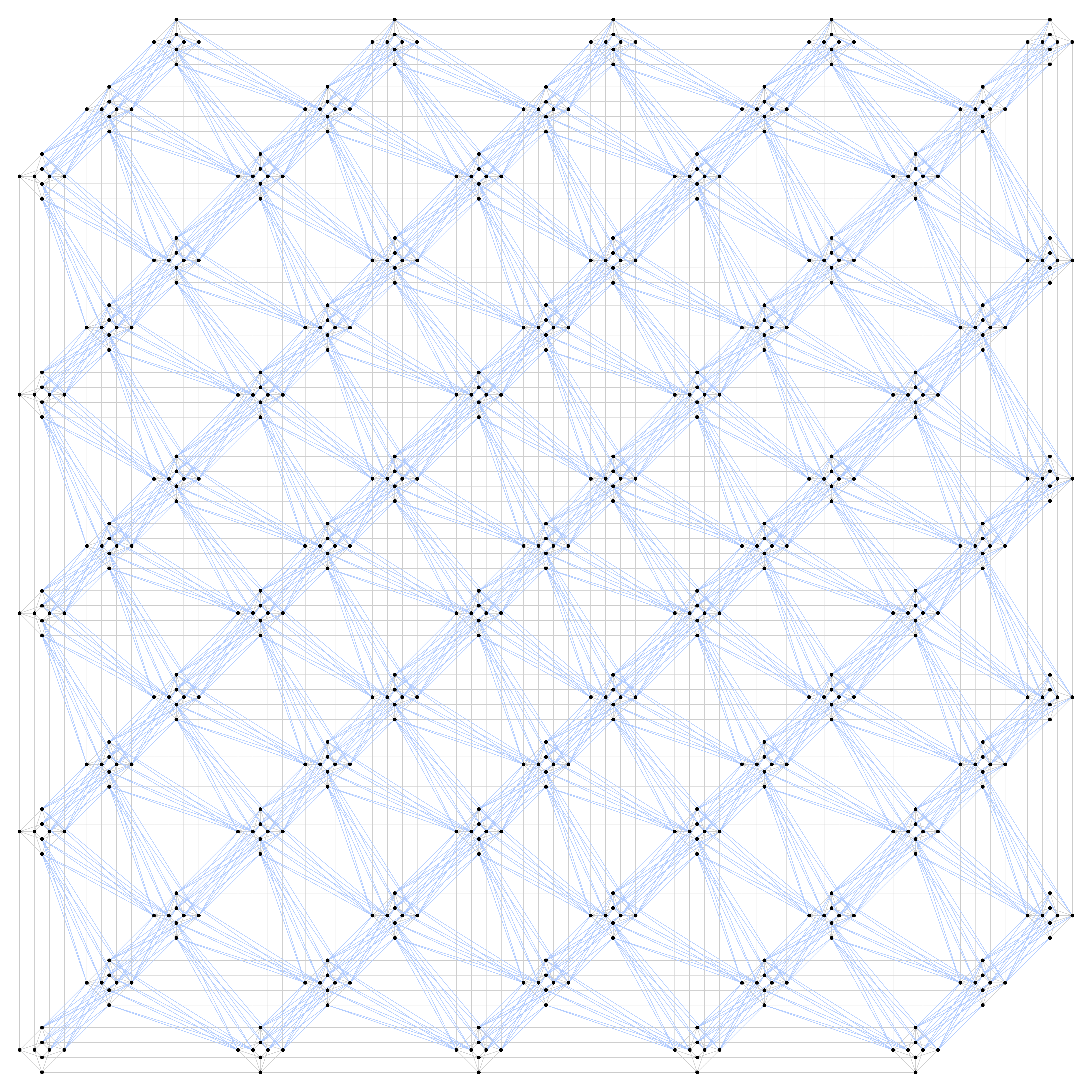}}~~~~~~~~~~

\vspace{10mm}

\subfloat[Triangle\vspace{2mm}
]{

\includegraphics[width=0.48\textwidth]{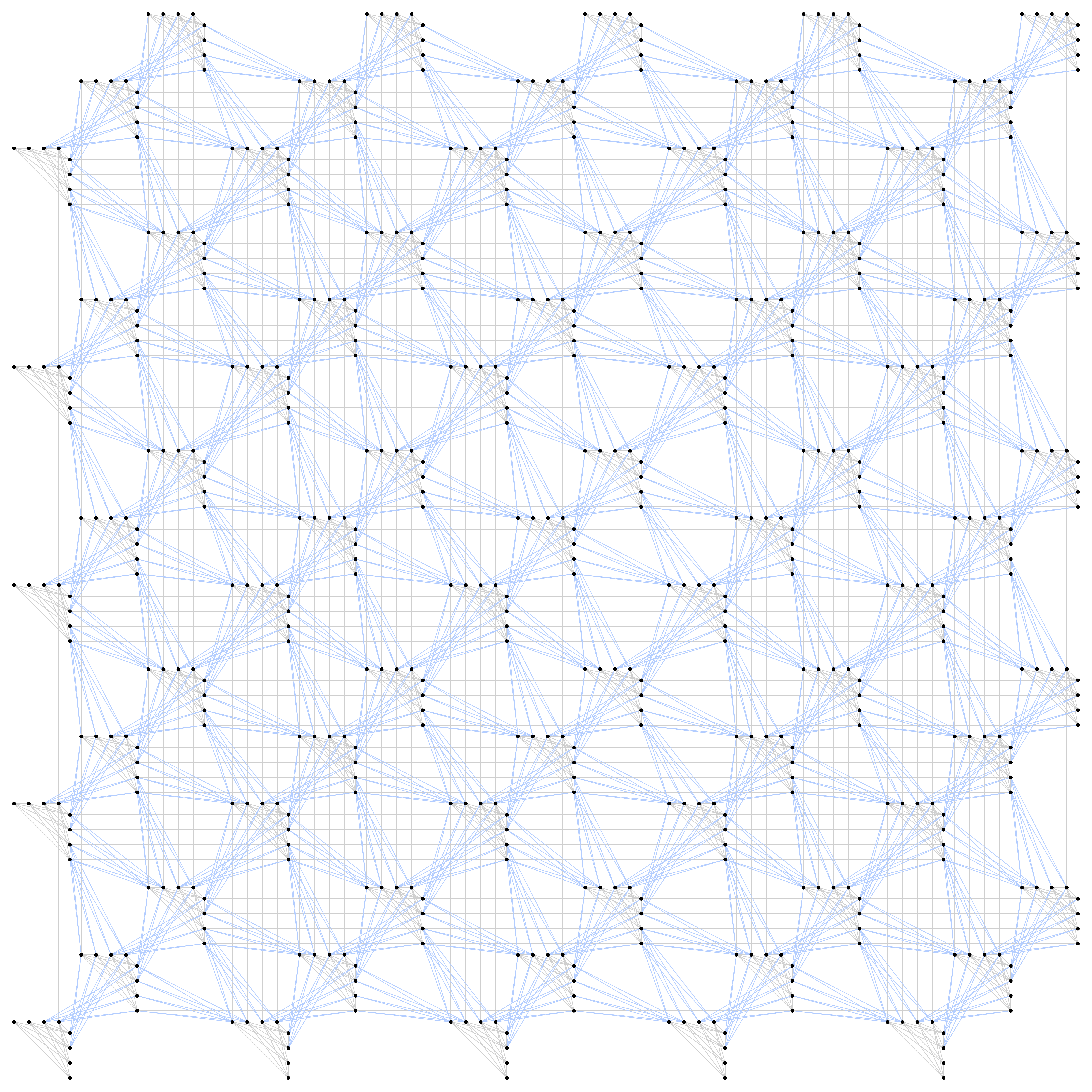}}~~~~\subfloat[Compressed\vspace{2mm}
]{

\includegraphics[width=0.48\textwidth]{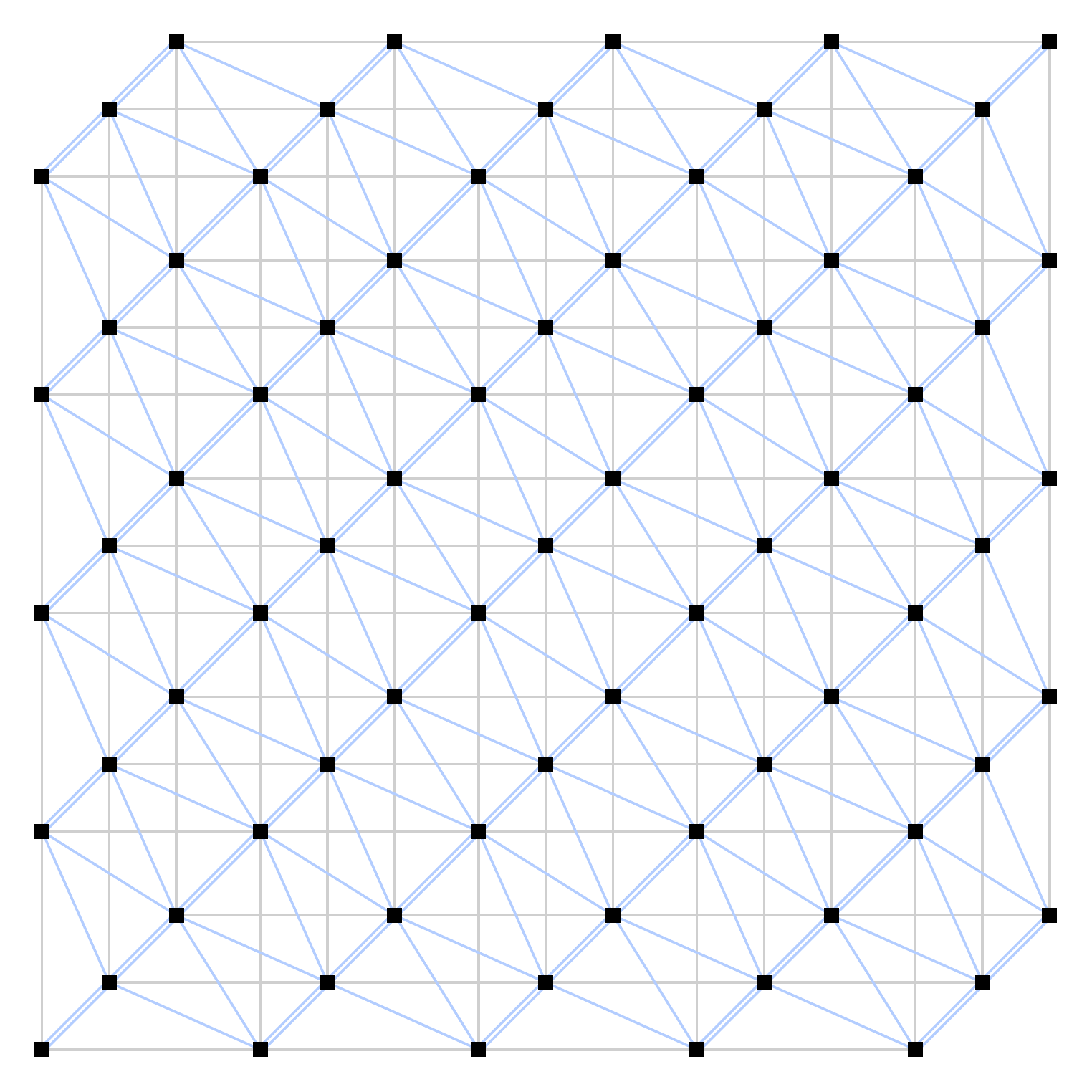}}
\end{figure}
\vfill{}
\par\end{center}

\begin{center}
\begin{figure}[H]
\vspace{10mm}
\caption{\centering$(X,Y,Z)=(5,5,3)$ \emph{tilted }lattice of Pegasus (with
all Pegasus-only edges either black or light blue).}
\vspace{10mm}

\subfloat[Tilted classic\vspace{2mm}
]{

\includegraphics[width=0.48\textwidth]{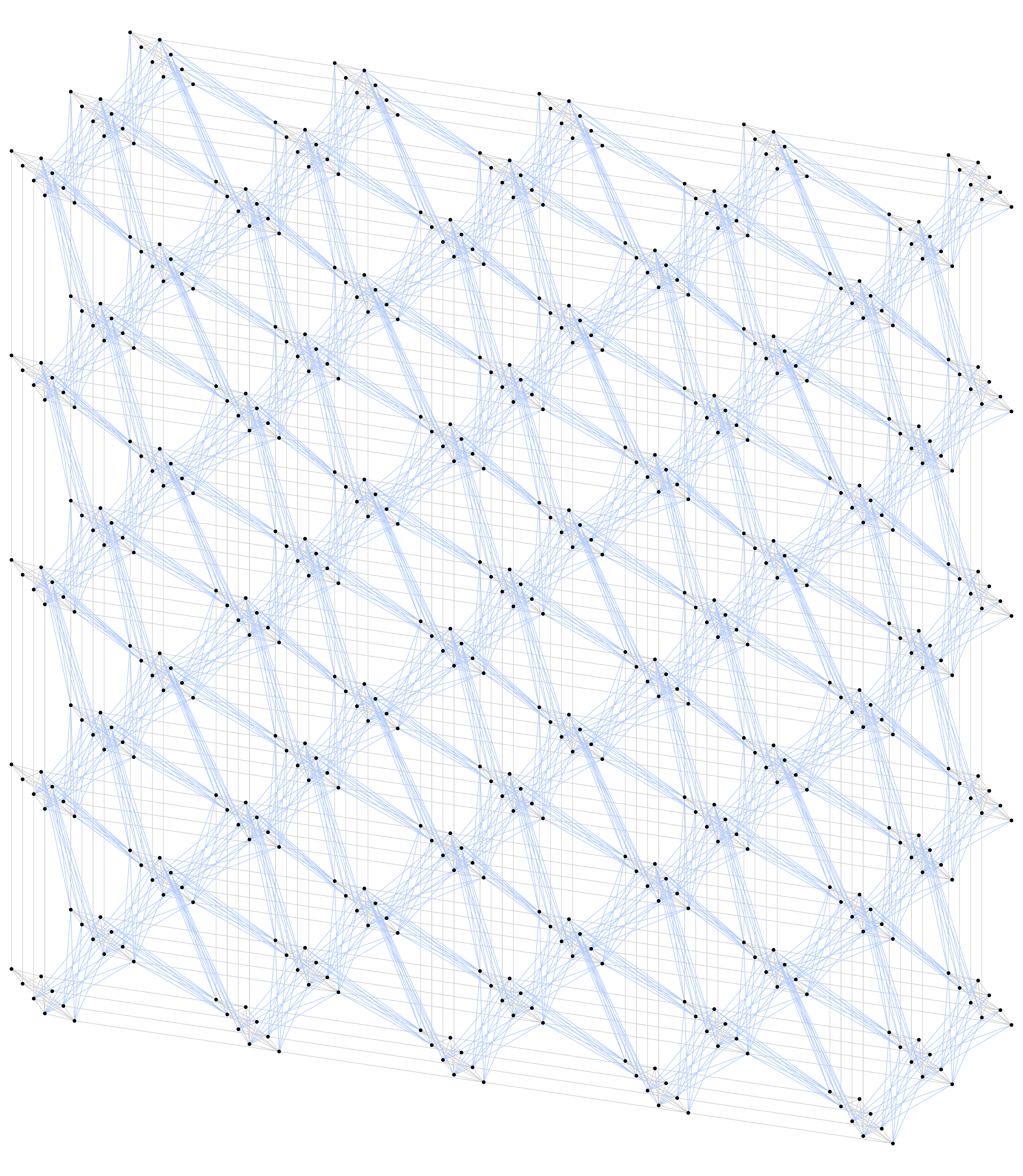}}~~~~~~~\subfloat[Diamond\vspace{2mm}
]{

\includegraphics[width=0.48\textwidth]{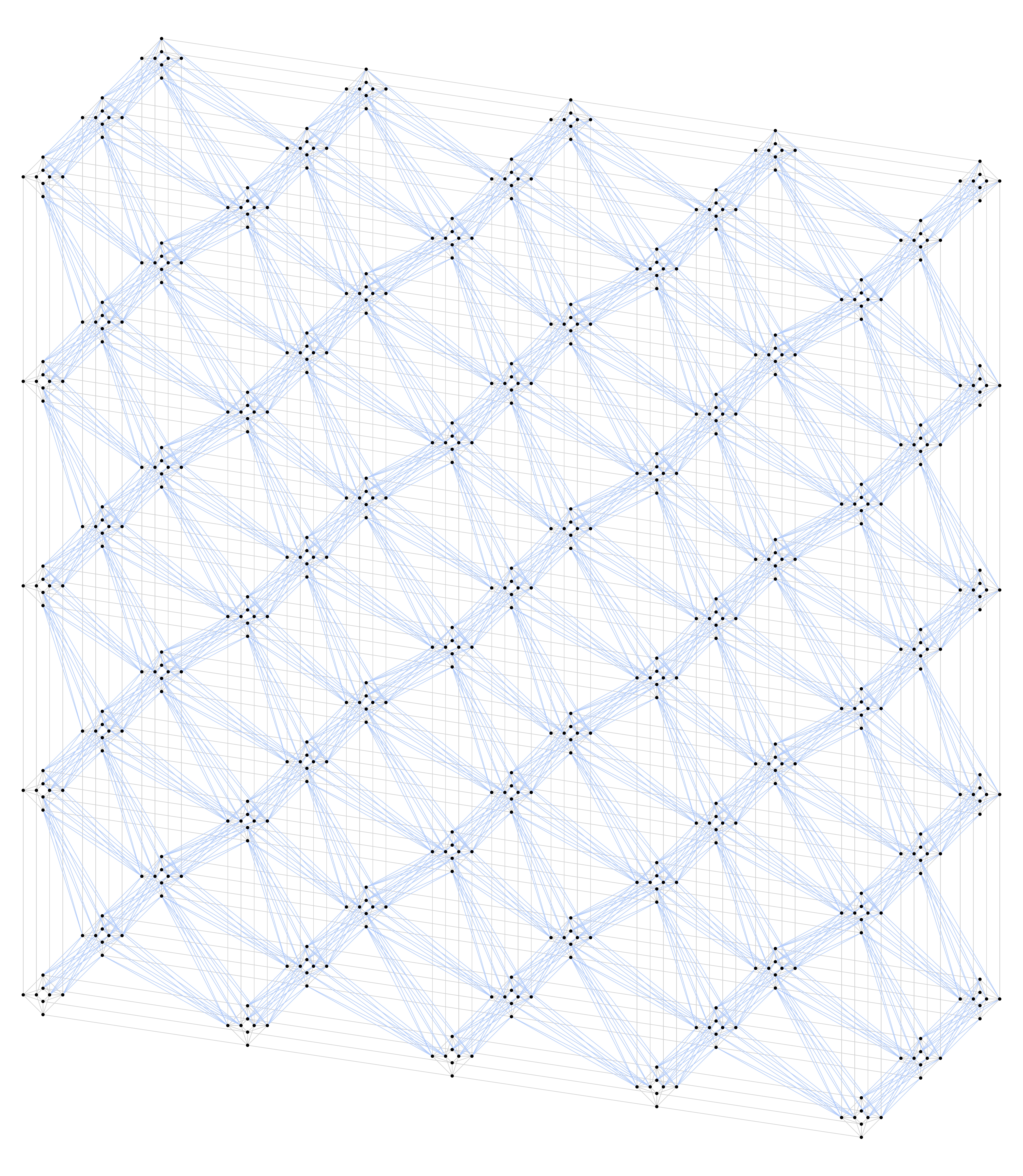}}~~~~~~~~~~\vspace{10mm}

\subfloat[Triangle\vspace{2mm}
]{

\includegraphics[width=0.48\textwidth]{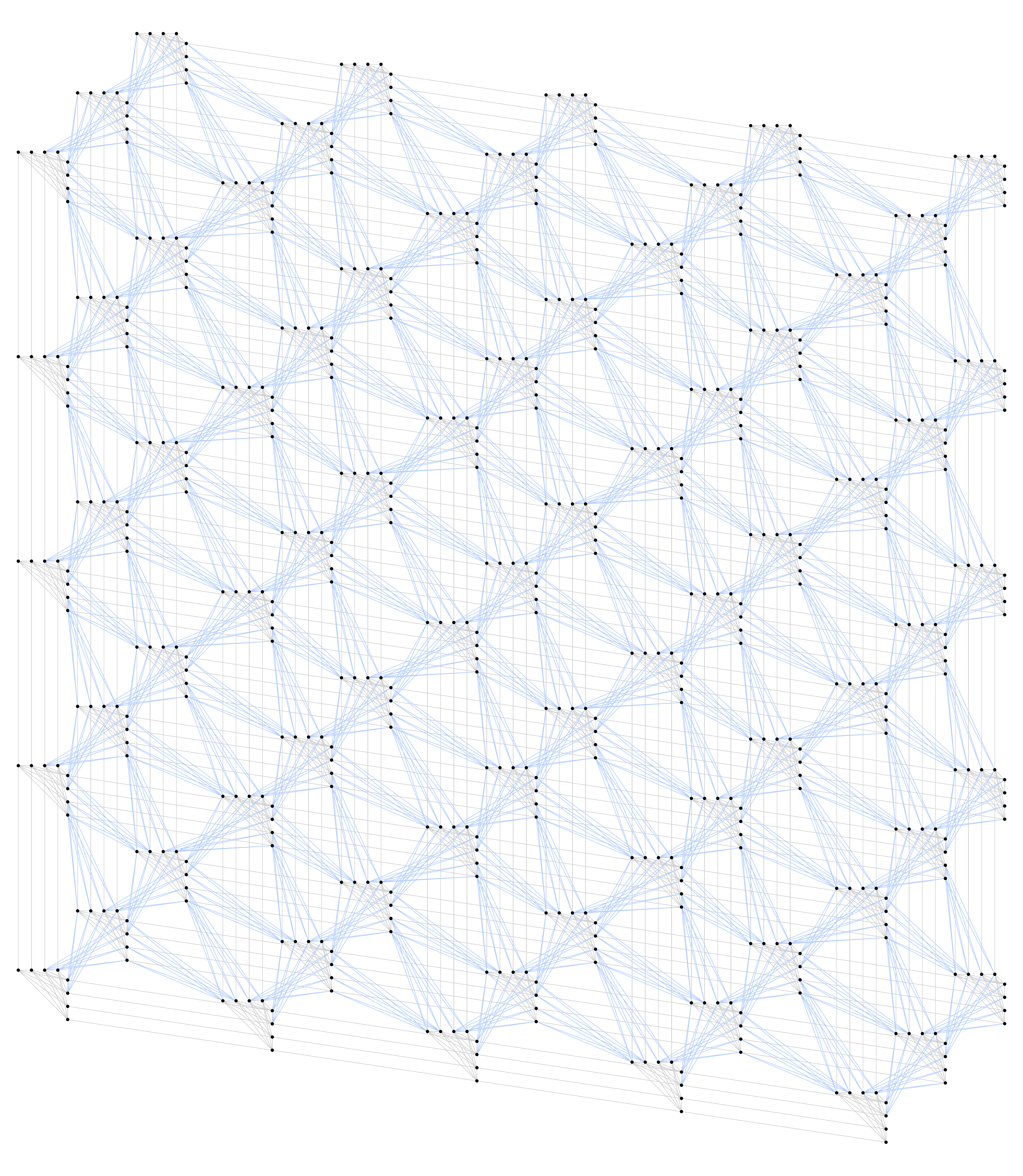}}~~~~~\subfloat[Compressed\vspace{2mm}
]{

\includegraphics[width=0.48\textwidth]{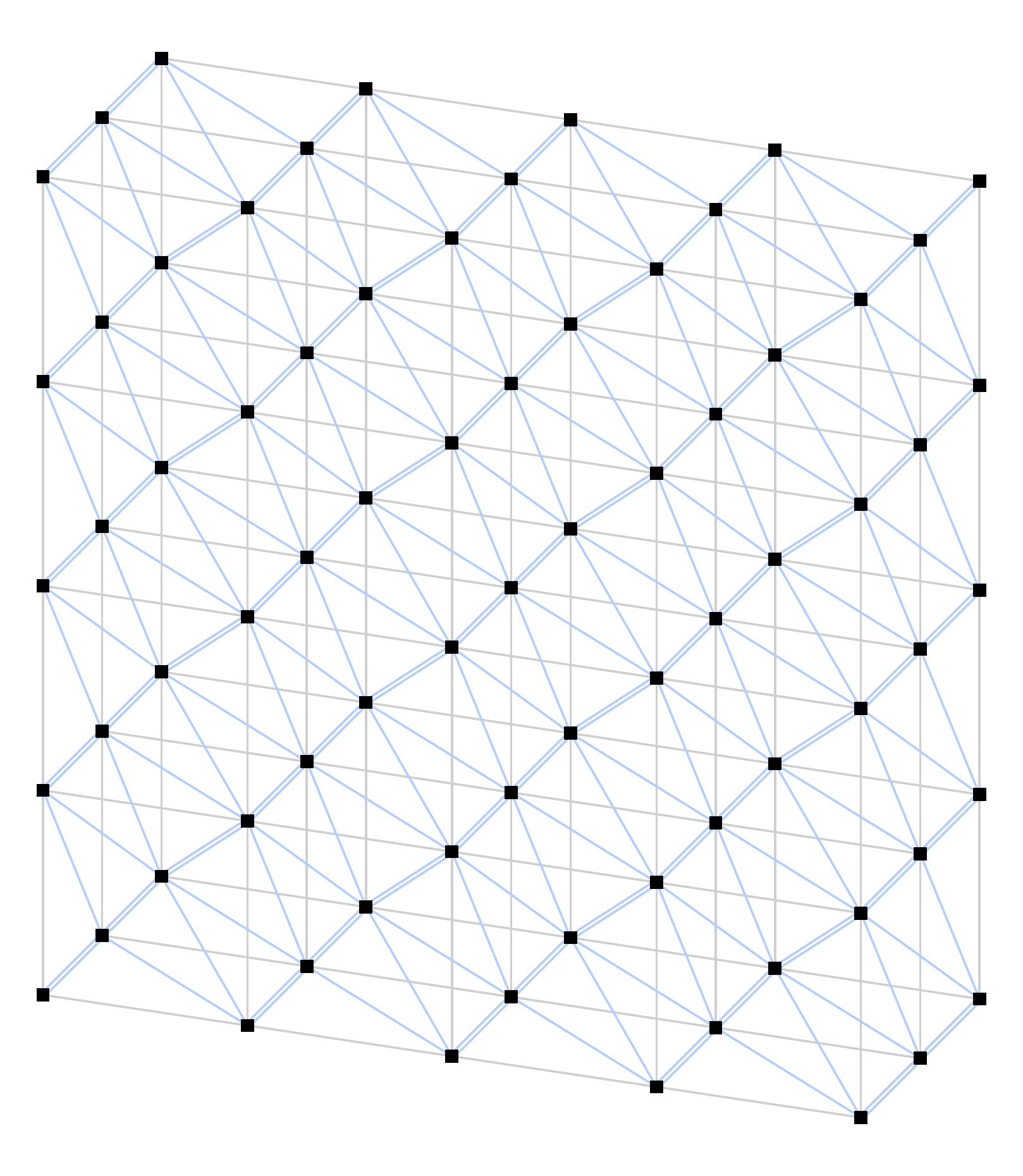}}
\end{figure}
\par\end{center}

\end{center}
\end{document}